\documentclass[final,3p,times]{elsarticle}

\usepackage{amssymb}
\usepackage{xcolor}
\usepackage{multirow}
\usepackage[percent]{overpic}
\usepackage{tabularx}
\newcommand{\probP}{\text{I\kern-0.15em P}}
\usepackage{multirow}
\usepackage[table]{xcolor}
\usepackage{colortbl}

\usepackage{graphicx} 
\usepackage{makecell}

\usepackage[table]{xcolor}
\usepackage{colortbl}
\usepackage{pgf}

\usepackage{amsmath}
\usepackage{xcolor}
\usepackage{colortbl}
\usepackage{xcolor}
\usepackage[
    colorlinks=true,
    citecolor=green,
    linkcolor=black,
    urlcolor=red
]{hyperref}
\usepackage{tikz}
\usepackage[ruled,vlined,linesnumbered]{algorithm2e}
\usetikzlibrary{trees, mindmap, shadows}
\usepackage{graphicx}
\usepackage{subcaption}
\usepackage{forest}
\usepackage{xcolor}
\usepackage{adjustbox}
\usepackage{booktabs}
\usetikzlibrary{positioning, shapes.geometric, arrows.meta}
\usepackage[most]{tcolorbox}

\journal{Computer Speech And Language}

\begin{document}

\begin{frontmatter}



\title{Phonetically Explainable Speech Deepfake Detection}




\author{Manasi Chhibber\corref{cor1}}\ead{manasi.chhibber@uef.fi}

\author{Jagabandhu Mishra}

\author{Tomi H. Kinnunen}

\cortext[cor1]{Corresponding author}

\affiliation{organization={School of Computing},
            addressline={University of Eastern Finland}, 
            city={Joensuu},
            postcode={FI-80101}, 
            country={Finland}}
            
\begin{abstract}
Speech deepfake detection, the task of determining whether an utterance originates from a genuine human speaker or a generative model, has predominantly been treated as an opaque classification task. In this setting, utterance-level scores typically emerge from aggregating acoustic evidence through mechanisms such as temporal averaging or learned attention, which treat all temporal frames as equally informative. This can be problematic, since different phonetic categories carry substantially different amounts of discriminative information. To address this limitation, we propose a phoneme-guided cross-attention framework that transforms detection into an interpretable, phonetically grounded process. Under three explicitly asserted probabilistic assumptions (sufficiency of the phonetic estimate, acoustic dominance over phonetic-estimator errors, and uninformative phonetic priors), we factorize the spoofing posterior $P(\text{spoofed}\mid X, W)$, conditioned on the acoustic representation $X$ and the phonetic posteriorgram $W$. The resulting factorization can be written in a short, intuitive form $P(\text{spoofed} \mid X, W) = \sum_{i=1}^{M} w_i \cdot P(\text{spoofed} \mid X, Z = z_i)$, where $M$ denotes the number of phonetic classes, each $P(\text{spoofed} \mid X, Z = z_i)$ is the spoofing probability for the $i$-th phonetic class $z_i$ conditioned on the acoustic representation $X$, and each $w_i$ is the prevalence of phonetic class $z_i$ in the utterance. This formulation reveals that discriminative power stems from per-phone class-conditional acoustic divergence, rather than from phoneme occurrence frequency. Our transformer-based architecture instantiates this decomposition through a cross-attention block in which phonetic anchors (extracted through a phone posteriorgram extractor) act as queries to selectively probe information in acoustic keys and values, with softmax-normalized pooling supplying explicit phone-presence weights. Unlike prior approaches that rely almost exclusively on post-hoc explainability methods, our framework offers phonetic-explainability-by-design. We evaluate the framework across three datasets of varied complexity: a controlled same-speaker, same-text LJSpeech-derived corpus, the standard ASVspoof 2019 LA benchmark, and the large-scale ASVspoof 5 Track 1 representative of state-of-the-art deepfakes and in-the-wild conditions. Across all three datasets, per-phone importance rankings consistently reveal that discriminative power concentrates on articulatory categories that generative models struggle to reproduce faithfully. Specifically, stops, fricatives, affricates, nasals, and silence-boundary closures are ranked most discriminative. Periodic, formant-driven vowels and semivowels, in turn, receive lower importance. A targeted phoneme-group ablation on ASVspoof 2019 LA where separate detection model is trained on each phone group confirms this importance ordering. Beyond competitive detection performance, a key feature of our model is structural interpretability, with each decision yielding an inspectable per-group breakdown of how each articulatory category contributed to the final verdict.
\end{abstract}




\end{frontmatter}


\section{Introduction}

Recent advances in deep generative modeling have made speech synthesis highly realistic, enabling text-to-speech (TTS) and voice conversion (VC) systems to convincingly reproduce synthetic voices from minimal training data~\cite{van2016wavenet,wang2023neural,tan2021survey}. While this technology offers many beneficial applications, the past decade has also seen a steady rise in \emph{speech deepfakes}---synthetic audio that, depending on the intended target, may aim to sound naturally human (when intended to deceive human listeners) or simply to carry the targeted person's speaker characteristics (when intended to spoof automatic systems). Such synthetic media have already been misused in real-world cases~\cite{robins_early_2024,brightside_deepfake_ceo} of impersonation, social engineering fraud, defamation, and circumvention of automatic speaker verification (ASV) systems \cite{bbc_cloned_voice_2024}. In response, for more than a decade already, the research community has actively developed \emph{anti-spoofing systems}, also known as \emph{countermeasures} (CMs). The primary aim of these systems is to analyze a given utterance and determine whether it originates from a genuine human speaker, or a generative model~\cite{wu2014voice,sanchez2015toward}.

Modern anti-spoofing systems have become highly effective at this binary real-fake classification task. The most robust systems are typically built upon self-supervised acoustic backbones, such as Wav2Vec~2.0~\cite{babu2021xls}, and employ end-to-end architectures like RawNet2~\cite{tak2021end} and AASIST~\cite{jung2022aasist}, with recent SSL-based detectors continuing to push the field forward through architectural innovations such as wavelet-domain sparse prompt tuning~\cite{xuan2026wavesp}, compact backbone studies~\cite{kulkarni2026compact}, and post-training adaptation~\cite{ge2025post}. Despite providing reasonable detection performance, these models remain fundamentally opaque. They process an entire utterance to produce a single scalar score, revealing very little about \emph{why} that specific (soft) decision was made. For instance, they neither elaborate on which specific temporal segments contain the spoofing traces, nor isolate the low-level acoustic features that triggered the model's prediction. As deployment scenarios expand beyond automated authentication to include forensic audits, legal investigations, and human-in-the-loop decision-making, this lack of decision transparency becomes a critical limitation. 

Recent literature argues that detectors must be not only accurate but also explicitly \emph{interpretable} at both linguistic and articulatory levels~\cite{khanjani2024aldas,khanjani2024investigating,yang2025forensic,fullwood2026mind,zhang2025phoneme}. Existing off-the-shelf approaches adopted from explainable artificial intelligence (XAI)~\cite{dovsilovic2018explainable} aim to explain detection decisions typically rely on post-hoc methods such as gradient-based attribution (e.g., Grad-CAM~\cite{liu24m_interspeech,gupta24b_interspeech}) or feature-importance analysis (e.g., SHAP-based attribution~\cite{ge2022explaining}). While these methods can offer useful insight into model behavior, they operate at a representational level that is decoupled from the architecture's internal computation, providing explanations that depend on the analyst's choice of probing technique rather than on the model's structural design. The present work takes a fundamentally different approach, building explainability into the architecture itself, with per-phone interpretability emerging as a structural property of the detection decision.

\begin{figure}[h!]
\centering
\begin{tikzpicture}[
    title/.style={
        draw=black,
        thick,
        fill=white,
        font=\bfseries\large,
        minimum width=4cm,
        minimum height=1.2cm,
        rounded corners=4pt,
        align=center
    },
    cat/.style={
        draw=black,
        thick,
        fill=white,
        font=\bfseries,
        minimum width=4cm,
        minimum height=1.5cm,
        rounded corners=4pt,
        align=center
    },
    itemtext/.style={
        font=\normalsize,
        align=center,
        text width=4cm
    }
]

\node[title] (root) at (0,0) {Speech Artifacts};

\node[cat] (c1) at (-5,-1.8)
{Signal Level\\(Global)};

\node[cat] (c2) at (0,-1.8)
{Segmental\\(Phonetic)};

\node[cat] (c3) at (5,-1.8)
{Suprasegmental\\(Prosodic)};

\draw[thick] (root.south) -- (c1.north);
\draw[thick] (root.south) -- (c2.north);
\draw[thick] (root.south) -- (c3.north);

\node[itemtext] at (-5,-3.3) {
Spectral Envelope\\
Phase Issues\\
Silence/Noise Floor
};

\node[itemtext] at (0,-3.3) {
Vowel Formants\\
Fricative Noise\\
Plosive Bursts
};

\node[itemtext] at (5,-3.3) {
Co-articulation\\
Breathing\\
Pitch Anomalies
};

\end{tikzpicture}

\caption{
Taxonomy of speech artifacts across three linguistic levels:
global \emph{signal-level} artifacts that arise from low-level acoustic
modeling choices in synthesis pipelines (spectral envelope smoothing,
phase discontinuities, and silence-floor anomalies);
\emph{segmental} (phonetic-level) artifacts that emerge in specific
speech sounds where generative models struggle to reproduce articulatory
mechanisms (vowel formant errors, fricative noise spectral irregularities,
and plosive burst smearing); and \emph{suprasegmental}
(prosodic-level) artifacts that affect the structural and intonational
features of an utterance (co-articulatory transitions, breathing patterns,
and pitch contour anomalies). The present work focuses on the
segmental level, which provides a granular and physically grounded
basis for explicit phonetic interpretability in speech deepfake detection.
}
\label{fig:neural_artifacts_columns}

\end{figure}
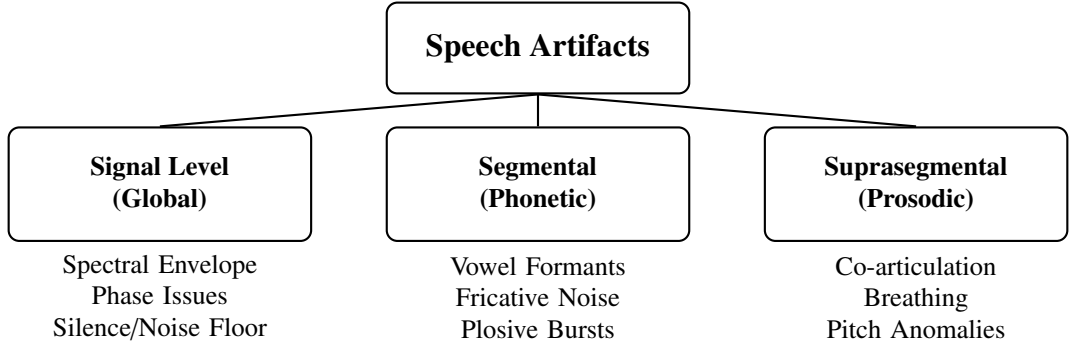

Achieving phonetically-guided interpretability requires a more careful look at the speech signal itself, since the processing artifacts exploitable by deepfake detectors are not uniformly distributed across it. As summarized in Figure~\ref{fig:neural_artifacts_columns}, these artifacts span three distinct levels. At the \emph{signal level}, generative pipelines leave global traces such as spectral-envelope over-smoothing, phase discontinuities, and silence-floor anomalies, which have motivated nearly two decades of low-level signal analysis~\cite{saito2017statistical,chappell2002comparison,todisco2017constant,hanilci2016spoofing}. At the \emph{suprasegmental} or \emph{prosodic} level, deviations in pitch contours, breathing patterns, unnatural pauses, and co-articulatory dynamics carry strong discriminative information~\cite{khanjani2024aldas,khanjani2024investigating,doan2023bts}. Between these two extremes lies the \emph{segmental} or \emph{phonetic} level. Here, modern TTS or VC systems aim to reproduce faithfully the specific spectral and temporal characteristics of individual speech segments, such as formant structures for vowels, transient bursts for plosives, turbulent broadband noise for fricatives, and anti-resonance behavior for nasals. Consequently, this is where synthesizers frequently fail in systematic, articulatorily explicable ways~\cite{pons2021upsampling,pandey2022production,fullwood2026mind,yang2025forensic}. The present work operates strictly at this segmental level. This choice is well justified. The segmental level offers a granular, physically grounded basis that allows detection decisions to be explicitly explained and audited, while remaining tightly coupled to the acoustic evidence upon which deep learning models already rely. The segmental level is also relevant to forensic voice comparison practice, where analysts have long isolated specific phonetic segments for targeted acoustic analysis~\cite{yang2025forensic,dar2025impact}. Therefore, early adaption of deepfake detection practices compatible with already-established forensic analysis approaches is expected to increase the likelihood that the resulting analysis methodologies will be adopted by forensic laboratories.

To understand why the segmental level offers this analytical handle, it is helpful to recall what the segments of speech actually are. Speech is a continuous acoustic signal, and phonetics provides the standard descriptive framework that organizes this continuous flow into a small inventory of distinctive sound categories known as, \emph{phones}. These include vowels, plosives, fricatives, nasals, and the brief silences and closures that separate them. As illustrated in Figure~\ref{fig:neural_artifacts_columns}, and supported by a number of recent phonetic analyses of synthetic speech~\cite{fullwood2026mind,yang2025forensic,dar2025impact}, these classes are not equally easy for a generator to reproduce; the periodic spectra of vowels are well suited to frame-based vocoders~\cite{kong2020hifi,morrison2021chunked}, whereas the transient bursts of plosives~\cite{pandey2022production}, the broadband turbulence of fricatives~\cite{pons2021upsampling}, and the anti-resonance behavior of nasals~\cite{fullwood2026mind} are notoriously harder to synthesize faithfully. Despite this, most existing deep anti-spoofing systems lack an explicit mechanism for inspecting each phonetic segment separately and assessing its authenticity. Instead, they treat the speech signal as a single acoustic stream, reducing it through pooling mechanisms such as temporal averaging, statistics pooling, or learned attention~\cite{snyder2018x,okabe2018attentive,tak2021end,jung2022aasist}. Even learned attention, which directs context-dependent focus to specific parts of the sequence, remains phone-agnostic and therefore does not align its decisions to the underlying phonetic structure of the utterance. Consequently, the vast majority of anti-spoofing systems produce a single utterance-level score that says nothing about which segments drove the decision. The segmental structure of speech, despite being the level at which generators differ most visibly from human speech production, remains, at best, an emergent property of the learned representations rather than a direct object of spoofing analysis. We argue that closing this gap by making the segmental phonetic structure an architecturally addressable component of the detection decision is the natural step forward, and it is the step that our present work takes.

\begin{figure}[h!]
    \centering
    \includegraphics[width=\linewidth]{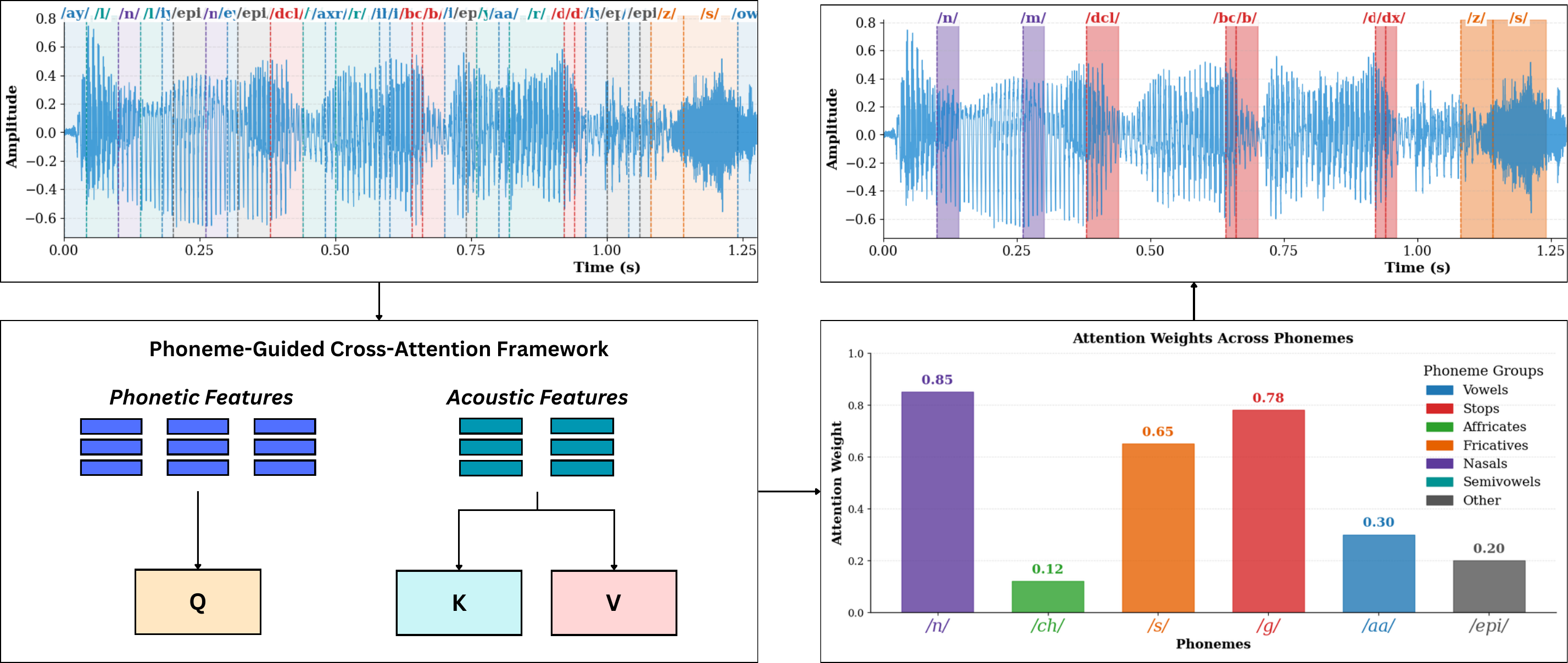}
    \caption{Overview of the proposed phoneme-guided cross-attention framework. The input speech waveform (top left) is annotated with its underlying phonetic segments. A cross-attention module fuses two parallel streams of information, with phonetic posteriorgrams supplying the queries and self-supervised acoustic embeddings supplying the keys and values. The resulting per-phone weights are visualized as a ranking across articulatory groups (color-coded), exposing which phonetic categories the model attended to most strongly. Projecting these weights back onto the original waveform (bottom) highlights the discriminative regions, allowing the detector to narrow its focus from the whole utterance to the phonetic segments that carry the most informative spoofing cues.}
     \label{fig:intro}
\end{figure}

In particular, we propose a phoneme-guided cross-attention framework that makes the per-phone structure of the detection decision \emph{phonetically explicit}. The conceptual core of the framework, illustrated in Figure~\ref{fig:intro}, is to combine a self-supervised acoustic stream (XLS-R~\cite{babu2021xls}) with a phonetic stream provided by frame-level phoneme posteriorgrams (PPGs)~\cite{hazen2009query} through a cross-attention mechanism in which the phonetic stream supplies the queries, which act as \emph{phonetic anchors} that interrogate the acoustic representation, while the acoustic stream supplies the keys and values. By construction, the resulting attention output contains one row per phonetic class, and a subsequent weighted-pooling stage produces an explicit, softmax-normalized set of per-phone weights that combine these rows into the final score. The architecture thus exposes, for every utterance, which phonetic classes the model attended to most strongly when forming its decision. As Figure~\ref{fig:intro} illustrates, this transforms deepfake detection from an opaque, utterance-wide judgment into a structured analysis. The signal regions corresponding to the most heavily weighted phonetic classes become visible at a glance. Rather than looking at the whole signal, our proposed model effectively narrows its focus onto the phonetic segments where the most informative spoofing cues are concentrated.

The contributions of our work are six-fold. First, given the widely scattered literature on phonetical aspects of deepfakes, we provide a self-contained and accessible review of the phonetic foundations relevant to deepfake detection, including the articulatory grouping of speech sounds and the typology of synthesis failure modes. This part of the work is to make the material more accessible to readers less familiar with the associated phonetical terminology (Sections~\ref{sec:rw_paralinguistic}--\ref{section_3_1}). Second, we formalize the deepfake detection problem through a probabilistic factorization of the spoofing posterior into a weighted sum of phone-conditional evidence terms modulated by phone-presence weights. Importantly, we explify three probabilistic assumptions (the interested reader is invited to peek at the colored box in Section \ref{factorization_and_assumptions}), from which the suggested phonetically-decomposed detection score follows. Third, we propose a phoneme-guided cross-attention architecture that realizes both quantities of this factorization explicitly, with a hybrid query construction that combines learned phonetic prototypes with utterance-specific context, and a learned weighted-pooling backend that exposes per-phone weights at inference time. Fourth, we conduct an extensive progressive validation across three datasets of varied complexity: a controlled, single-speaker corpus (LJSpeech (TTS-derived)) generated in-house using Tacotron~2~\cite{shen2018natural}  as the acoustic model and Parallel WaveGAN~\cite{yamamoto2020parallel} as the neural vocoder; a standard multi-speaker benchmark with unseen attacks (ASVspoof~2019 LA~\cite{wang2020asvspoof}); and a large-scale in-the-wild evaluation with neural-codec and adversarial conditions (ASVspoof~5 Track~1~\cite{delgado2024asvspoof}). Fifth, we run a targeted articulatory-group ablation on ASVspoof~2019 LA, which independently confirms that under a controlled experimental protocol, the relative ordering of articulatory categories already implied by the model's learned attention weights is preserved. Sixth, we present interpretable score-decomposition case studies in which the per-utterance contribution of each articulatory category to the final decision is rendered transparently, demonstrating the practical utility of the framework as a tool for understanding generative-model weaknesses, not merely a detector of them. To sum up, the main scientific and technical aim of the present work is to define a rigorous reference framework for phonetically-explainable speech deepfake detection. A key contribution is that the framework introduces \textbf{phonetic-explainability-by-design}, with per-phone interpretability built into the architecture itself rather than offered as a post-hoc analytical add-on. Our work sits at the intersection of phonetics, probability theory, and state-of-the-art foundational-model–based deep learning.

\section{Related Work: The Phonetic Basis of Spoofing Artifacts}
\label{related_works}

Whereas the majority of speech deepfake detection research has approached the problem from generic machine-learning and signal-processing perspectives, a recent, emerging line of work has begun to ground detection more explicitly in the phonetic structure of speech. In this section, we provide a brief chronological summary of how phonetic information has been incorporated into anti-spoofing systems.

\subsection{From Handcrafted Acoustic Features to Bottleneck-Feature Beginnings}
\label{sec:rw_history}

Early anti-spoofing research was dominated by signal-processing approaches targeting the artifacts of statistical parametric speech synthesis (SPSS) and unit-selection synthesis~\cite{wu2014voice,sanchez2015toward}, under the assumption that synthetic speech exhibited over-smoothing in the spectral envelope~\cite{saito2017statistical} and phase discontinuities detectable through short-term spectral analysis~\cite{chappell2002comparison,hanilci2016spoofing}. Within this family, time-frequency analysis methods based on the short-term Fourier transform~\cite{sahidullah2015comparison} and the constant-$Q$ representation~\cite{todisco2017constant} produced the dominant feature representations of the era. These features were typically modeled using Gaussian mixture models (GMMs), an approach that is fundamentally ``phonetically blind'' in the sense that it models only the global distribution of acoustic features and discards both temporal sequence and linguistic context.

Recognizing the limitations of purely acoustic modeling, early attempts incorporated phonetic context using \emph{bottleneck features} (BNFs), a representation originally designed for applications such as automatic speech recognition (ASR)~\cite{grezl2007probabilistic,yilmaz2019articulatory}. An early study~\cite{alam2016spoofing} on the ASVspoof~2015 corpus found BNFs to be useful especially in detecting unit-selection-based spoofing attacks, providing an early hint at the usefulness of phonetic information for anti-spoofing. Nonetheless, being based on the now-outdated HMM--DNN ASR architectures of the era and relying on shallow fusion mechanisms such as feature concatenation, this generation of BNF-based countermeasures can be argued to possess limited capacity to exploit the phonetic stream in a structured way.

\subsection{Paralinguistics- and Prosody-Inspired Approaches}
\label{sec:rw_paralinguistic}

A second wave of work engages with non-phonetic but linguistically meaningful events in the speech signal. The BTS-E framework~\cite{doan2023bts}, for instance, models the breathing, talking, and silence states of an utterance, recognizing that natural phenomena such as breathing carry their own acoustic signatures that generative models struggle to reproduce faithfully~\cite{sun2023exposing}. Such non-phonetic events span a broader typology including physiological sounds, affective vocalizations, disfluencies, and paralinguistic voice qualities, and segment-aware detectors can in principle exploit failures in any of these. A parallel line of work targets higher-level prosodic cues directly: hand-labeled features such as breath intake, pitch anomalies, and unnatural pauses~\cite{khanjani2024aldas} have been analyzed through causal discovery algorithms~\cite{khanjani2024investigating}, identifying pitch- and audio quality-related anomalies as being the causally most relevant factors for spoofing detection. These paralinguistics- and prosody-inspired approaches operate at a level largely orthogonal to the phonetic content of the utterance and are therefore complementary to the phonetic perspective adopted in the present work, which centers the analysis on the speech sounds themselves.

\subsection{Phonetically-Aware Approaches} \label{sec:rw_phonetic}
A more explicitly phonetic strand of recent work pursues a \emph{divide and conquer} strategy that separates the analysis of what is being said (content) from how it is being produced (style/artifact). A few preliminary anti-spoofing studies have borrowed this idea to decouple linguistic or speaker identity from generation-style artifacts, but the direction remains largely underexplored, with early methods often failing to achieve clean separation without discarding useful forensic cues~\cite{wang2025generalize,li2025critical}.

\subsection{Phonetically-Informed Forensic Analysis of Synthetic Speech}

A complementary, more forensically oriented line of research applies rigorous phonetic analysis techniques to characterize the acoustic differences between bonafide and synthetic speech. Using automatic formant tracking together with phonetic time-alignment of the signal, a recent study~\cite{dar2025impact} indicates that deepfakes exhibit subtle deviations in the vowel space, either centralizing vowels or producing physically implausible formant bandwidths. Physically, formant bandwidths are related to energy damping in the vocal tract through tissue absorption, friction, and lip radiation~\cite{ladgfoged2001course}, with the lower formant bandwidths in natural speech being typically in the range 40--200~Hz; generative models often fail to replicate this damping, yielding bandwidths that are unnaturally narrow or physically implausible~\cite{dar2025impact}. Phone-class analyses further exposes consistent vocoder weaknesses: neural vocoders using transposed or dilated convolutions for upsampling introduce tonal and high-frequency artifacts such as spectral replicas and metallic coloration, particularly audible in the broadband noise of fricatives (e.g., \texttt{/s/}, \texttt{/f/})~\cite{pons2021upsampling}, while plosives tend to be over-smoothed by GAN-based vocoders such as HiFi-GAN into a muffled perception quality~\cite{pandey2022production}. Segmental features have been shown to consistently outperform global long-term statistics for detection~\cite{yang2025forensic}, supporting the premise that detection should be phoneme-specific. Extending this inquiry to contemporary architectures, a large-scale evaluation of 23 modern speech generators in~\cite{fullwood2026mind} showed that, while pitch contours are reproduced with high fidelity, the spectral traits of specific phone classes, notably nasals (e.g., \texttt{m}) and obstruents (e.g., \texttt{t}), remain poorly modeled, with errors traceable to the text-to-spectrogram stage \cite{klein2024source,zhu2022source} rather than downstream vocoding, revealing a persistent limitation in semantic-to-acoustic mapping. Table~\ref{tab:phoneme_failure_modes} summarizes the articulatory features, typical synthesis failure modes, and resulting acoustic detection cues for major phoneme classes.

\begin{table*}[h!]
\centering
\caption{Summary of phoneme-class-specific articulatory features, alongside typical synthesis failure modes and potential detection cues. The listed failure modes are grounded in established observations from recent speech synthesis and anti-spoofing literature, highlighting consistent vulnerabilities in generative models.}
\label{tab:phoneme_failure_modes}
\begin{tabular}{p{3cm} p{3.5cm} p{4.5cm} p{4.5cm}}
\hline
\textbf{Phoneme Class} & \textbf{Articulatory Feature} & \textbf{Typical Synthesis Failure Modes} & \textbf{Potential Detection Cues} \\
\hline
Plosives (e.g., \texttt{p}, \texttt{t}, \texttt{k}) &
Sudden burst of air (transient) &
Over-smoothing; lack of sharp onset; ``smearing'' of burst energy~\cite{pandey2022production} &
Reduced energy in high-frequency transient bands; slower rise time in the waveform envelope \\

Fricatives (e.g., \texttt{s}, \texttt{sh}) &
Turbulent airflow (noise) &
Tonal or metallic artifacts; checkerboard patterns from upsampling; bandwidth limiting~\cite{pons2021upsampling} &
Periodic structures in high-frequency spectrograms; anomalies in noise spectral density \\

Nasals (e.g., \texttt{m}, \texttt{n}) &
Velum lowering (anti-resonance) &
Incorrect spectral zeroes (anti-formants); leakage of oral resonance \cite{fullwood2026mind} &
Mismatch in spectral tilt and formant bandwidths; ``muffled'' quality due to smoothing \\

Vowels (e.g., \texttt{a}, \texttt{i}, \texttt{u}) &
Periodic vocal fold vibration &
Generally well modeled, but susceptible to style–linguistics mismatch (e.g., incorrect formant positioning for the target speaker)~\cite{dar2025impact} &
Inconsistencies between F0 contours and formant trajectories; formant bandwidth errors \\
\hline
\end{tabular}
\end{table*}

\subsection{Towards Architectures That Exploit Phonetic Structure}
\label{sec:rw_architectures}

A more recent line of work moves beyond analysis to embed phonetic structure directly into the detector architecture itself, producing what we term \emph{phonetically-aware architectures}. For instance,~\cite{zhang2025phoneme} introduces phoneme-level feature discrepancies through adaptive phoneme pooling: a pre-trained Wav2Vec~2.0 model segments the audio, and acoustic features within each segment are averaged into a single per-phone vector. These vectors are then processed by a \emph{graph attention network} (GAT)~\cite{velivckovic2017graph} to model temporal dependencies between successive phones. The authors reported that deepfakes often replicate the static spectral characteristics of individual sounds but fail to capture realistic transition dynamics arising from \emph{co-articulation}~\cite{daniloff1973defining}, that is, the natural overlap of articulatory movements caused by vocal-tract inertia. This insight aligns closely with the intuition behind our framework, albeit realized through a graph-based formulation rather than the cross-attention prototypes that we develop.

The difficulty of faithfully reproducing segmental and co-articulatory dynamics is further compounded when generators aim to simultaneously preserve linguistic content and transfer expressive prosody. For instance, recent work~\cite{nallaguntla2026phoneme} extended phonetic analysis to emotional voice conversion using WavLM embeddings and symmetric Kullback--Leibler divergence to measure how much each phonetic category's acoustic distribution shifts when emotion is imposed. The authors found out that emotional manipulation degrades speech non-uniformly: complex vowels, diphthongs, and fricatives are distorted substantially more than simpler single-vowel sounds (monophthongs), indicating that the more dynamic and articulatorily complex categories are the most vulnerable to emotional voice conversion.

Closely related to this line of phonetically-aware architecture work, another strand examines how the layered representations in self-supervised learning (SSL) models can themselves be exploited for phonetic-aware detection. The adoption of SSL models such as Wav2Vec~2.0, HuBERT, and XLS-R has substantially advanced the field; these models, trained on thousands of hours of multilingual speech, learn rich representations that encode phonetics, speaker identity, and many other linguistic and paralinguistic properties. Different layers of these transformer models encode different types of information. The study in~\cite{xiao2025layer} highlighted that standard feature fusion can lead to \emph{feature collapse}, where the model relies on only a few dominant layers and discards potentially discriminative signal from the others. The authors proposed training separate classifiers on each layer and subsequently fusing the resulting decisions, revealing that early-to-middle layers which are assumed to contain a mixture of acoustic and phonetic information, are often the most discriminative for deepfake detection.

XLS-R, the multilingually pre-trained variant of Wav2Vec 2.0, is particularly relevant for robust detection across languages~\cite{babu2021xls}. Deepfakes generated in one language often share artifact characteristics with those generated in another, provided the underlying vocoder is the same, and XLS-R provides a common phonetic space for such transfer, a capability leveraged in our proposed framework.


Taken together, the literature reveals a clear, emerging evolution: detection is moving from global spectral analysis to local, semantically grounded anomaly detection. Recent studies have begun to engage with phonetic structure in various ways~\cite{zhang2025phoneme,xiao2025layer,dar2025impact,yang2025forensic,fullwood2026mind}, but the field is still emerging, and to the best of our knowledge, no prior work makes per-phone interpretability the driving design principle of the detection architecture itself. The framework developed in the remainder of this paper builds on this trajectory: it uses PPGs as the structural interface between linguistic content and acoustic evidence and operationalizes a per-phone decomposition through a cross-attention architecture that explicitly exposes which articulatory categories drive each detection decision.

\section{Phonetic Background and Computational Extraction}
\label{section_3}

This section provides the minimal phonetic background to follow the rest of the paper, focusing on the articulatory categorization of speech sounds and on their computational extraction through phoneme posteriorgrams; readers seeking a more comprehensive treatment of acoustic phonetics and the production mechanisms of individual phone classes are referred to standard textbooks~\cite{ladgfoged2001course,stevens2000acoustic} and to recent surveys of distinctive-feature theory~\cite{jakobson1951preliminaries,hall2001distinctive}.

\subsection{Phonetic Foundations and Articulatory Groupings}
\label{section_3_1}

Human speech is a highly complex, but structured sequence of discrete linguistic units. A \textit{phoneme} is the smallest contrastive unit of sound in a given language: replacing one phoneme with another changes the identity or meaning of a word \cite{sapir2004language}. Phonemes are therefore abstract mental categories rather than physical events, conventionally described through bundles of articulatory attributes. These include features of voicing, nasality, and other dimensions of manner and place of articulation as formalized in distinctive feature theory~\cite{jakobson1951preliminaries,hall2001distinctive}. The physical realizations of phonemes are known as \emph{phones} (with context-dependent variants called \textit{allophones}), and in the acoustic signal they are usually referred to as \textit{segments}, in contrast to \textit{suprasegmental} phenomena such as intonation and stress.

The physical production of the segments is commonly modeled through the classical \textit{source--filter} framework \cite{fant1981source}, which separates the roles of excitation source (periodic vibration of the vocal folds for voiced sounds, turbulent or transient noise for unvoiced sounds) from the vocal tract acting as a time-varying filter shaped by constrictions of the tongue, lips, jaw, and velum. The resulting airflow dynamics such as turbulence, transient bursts, nasal coupling are not part of the biomechanical configuration itself but its acoustic consequence, giving each segment its characteristic spectral signature.

Whereas artificial speech generators such as text-to-speech and voice conversion need not faithfully mimic the physical speech production mechanism, they must approximate the intended phonetic content closely enough that listeners perceive the correct sequence of segments. This requires a deep generative model to reproduce not only the spectral and temporal characteristics of individual phones but also the dynamic (coarticulatory) transitions between them. Analyzing synthetic speech one segment at a time therefore provides a principled way to identify where, and on what kind of speech sound, a generator tends to fail.

To systematically examine phone-specific spoofing artifacts, we adopt the phonetic alphabet of the TIMIT corpus~\cite{garofolo1993timit}, which transcribes spoken English into 61 fine-grained phonetic labels (see Table~\ref{tab:phoneme_groups} for the full inventory). In addition to analyzing individual phones, we also consider their articulatory groupings into seven broad categories (\texttt{Vowels}, \texttt{Stops}, \texttt{Affricates}, \texttt{Fricatives}, \texttt{Nasals}, \texttt{Semivowels}, and \texttt{Other}) based on the hypothesis that the failure modes of synthesis systems follow a common logic within each category. Similar manner-based groupings have been widely used to analyze phone-level behavior in ASR and speech synthesis systems~\cite{lopes2011phone,fullwood2026mind}.

\begin{table*}[h!]
\centering
\caption{Articulatory grouping of the 61 TIMIT phonetic labels \cite{garofolo1993timit}, organized by manner of production. *For the purposes of acoustic grouping, \texttt{h\#} was treated as a fricative-like segment due to its turbulent spectral characteristics, although it is not universally classified as a canonical fricative in phonetics.}
\label{tab:phoneme_groups}

\begin{tabular}{p{2cm} p{13cm}}
\toprule
\textbf{Group} & \textbf{Phones} \\
\midrule

\textbf{Vowels} & 
\texttt{aa, ae, ah, ao, aw, ax, ax-h, axr, ay, eh, er, ey, ih, ix, iy, ow, oy, uh, uw, ux} \\

\textbf{Stops} & 
\texttt{b, d, g, p, t, k, dx, q, bcl, dcl, gcl, pcl, tcl, kcl} \\

\textbf{Affricates} & 
\texttt{ch, jh} \\

\textbf{Fricatives} & 
\texttt{dh, f, th, s, sh, v, z, zh, hh, hv, h\#*} \\

\textbf{Nasals} & 
\texttt{m, n, ng, em, en, eng, nx} \\

\textbf{Semivowels} & 
\texttt{l, r, w, y, el} \\

\textbf{Other} & 
\texttt{pau, epi} \\

\bottomrule
\end{tabular}
\end{table*}

\textbf{Vowels} constitute the most frequent and acoustically prominent group in human speech, characterized by a relatively open vocal tract and the continuous, quasi-periodic oscillation of the vocal folds~\cite{ladgfoged2001course}. This source–filter combination produces two distinct spectral signatures: a fine \textit{harmonic structure}, a series of equi-spaced spectral lines arising from the glottal source, and a \emph{spectral envelope} that varies slowly along the frequency axis. It is shaped by the vocal-tract resonances known as \textit{formants}~\cite{stevens2000acoustic}. Because modern neural vocoders excel at modeling periodic, highly correlated continuous signals, vowels are generally synthesized with high fidelity. Residual artifacts can nonetheless arise in the reconstructed phase spectrum, such as misaligned harmonic phases between adjacent frames, particularly in GAN-based and diffusion-based vocoders that operate primarily on spectral magnitude representations~\cite{kong2020hifi,morrison2021chunked}.

In stark contrast to vowels are the \textbf{stop} consonants (or plosives), which include sounds such as \texttt{p}, \texttt{t}, and \texttt{k}. The production of a stop consonant requires a complete mechanical closure of the vocal tract, followed by a brief buildup of intraoral pressure and a sudden, explosive release that generates a highly transient, broadband burst of acoustic energy \cite{ladgfoged2001course,stevens2000acoustic}. Critically, the release burst itself is extremely short, typically lasting only \(5\text{--}20\,\mathrm{ms}\) \cite{stevens2000acoustic}, which is comparable to or shorter than the analysis/synthesis frame sizes commonly used by neural vocoders (typically \(10\text{--}25\,\mathrm{ms}\) windows with \(5\text{--}10\,\mathrm{ms}\) hop size) \cite{kong2020hifi,tan2021survey}. These vocoders generate each output frame from a smoothed spectral summary (e.g., a mel-spectrogram) that averages acoustic content over a fixed analysis window, so transient events shorter than the window cannot be represented as sharp, localized bursts. As a result, stop consonants are frequently rendered as acoustically smeared or muffled~\cite{pandey2022production}.

Closely related to stops are the \textbf{affricates}, such as \texttt{ch} and \texttt{jh}, which begin with a complete articulatory closure but release into a turbulent, noise-like constriction. This dual-state nature requires the synthesis model to rapidly transition from silence to transient burst to sustained noise, making them highly complex to generate accurately. The next group, \textbf{fricatives}, which encompass sounds like \texttt{s}, \texttt{z}, and \texttt{sh}, are produced by forcing air through a narrow articulatory constriction, generating continuous high-frequency turbulent noise. Because the acoustic structure of fricatives is inherently stochastic rather than periodic, neural synthesis models often have difficulty reproducing their true noise characteristics, introducing unnatural periodic tones or a metallic "buzzing" artifact into the higher frequency bands \cite{pons2021upsampling}\cite{kong2020hifi}.

From the remaining groups, \textbf{nasals}, such as \texttt{m} and \texttt{n}, are produced by lowering the velum so that air resonates through the nasal cavity coupled with the closed oral cavity. This coupling introduces both resonances (formants, or \textit{poles}) and \textit{anti-resonances} (spectral \textit{zeros}) that must be jointly modeled; failure to do so results in a distinctly unnatural, synthetic timbre \cite{stevens2000acoustic}\cite{fullwood2026mind}. \textbf{Semivowels}, including liquids and glides like /l/, /r/, and /w/, share the periodic, voiced nature of vowels but function phonologically as consonants: like other consonants, they occupy the edges of a syllable rather than its centre, where vowels reside~\cite{ladgfoged2001course}. Their acoustic signatures are highly dynamic, defined chiefly by rapid, continuous formant transitions into adjacent vowels. Finally, the \textbf{other} category captures various forms of silence, including background pauses and the brief, natural closures that precede stop bursts. Although seemingly trivial, the near-silence produced by a human vocal tract is remarkably difficult for deep learning models to replicate without introducing a synthetic noise floor or low-amplitude periodic artifacts. This challenge has manifested concretely as a well-known dataset artifact: in ASVspoof~2019 for example, the silent regions of synthetic utterances contain literal zero-valued samples, while bonafide recordings preserve the low-level ambient background of the original recording environment, allowing detectors to exploit this shortcut \cite{geirhos2020shortcut} rather than learning genuine phonetic cues~\cite{chettri2020dataset,muller2021speech}. Using these seven articulatory groups, we provide a structured linguistic basis for evaluating exactly where and why deepfake models fail. Translating this linguistic structure into a usable signal for detection, however, requires a continuous, frame-level estimate of which articulatory class is active at each moment in the audio. We introduce such an estimator next, before formalizing how it is integrated into our detection framework.

\subsection{Phonetic Posteriorgram: Estimating the Latent State}
\label{section_3_2}

\begin{figure}[h!]
    \centering
    \includegraphics[width=\linewidth]{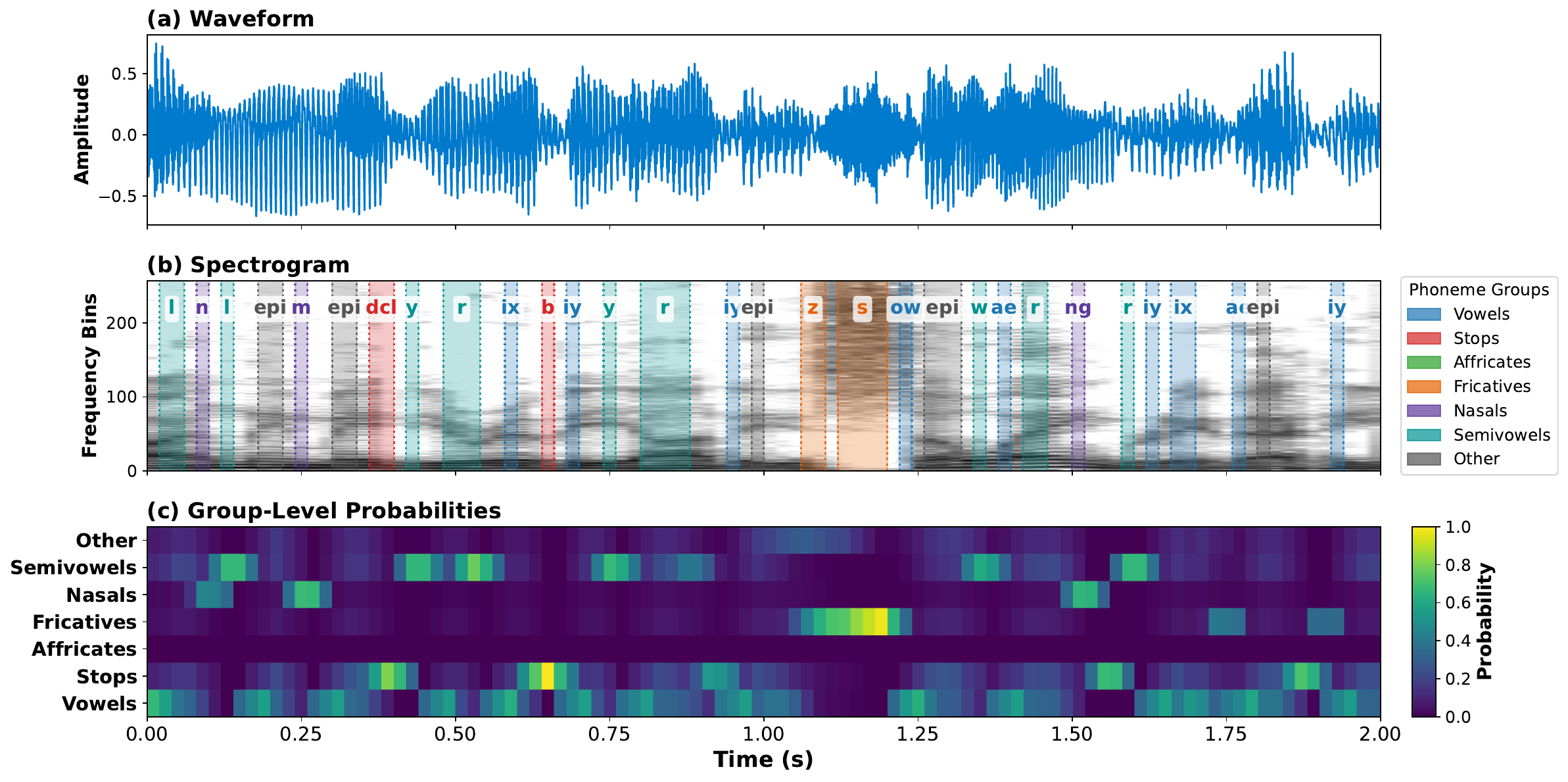}
    \caption{Visualization of the temporal alignment and probabilistic estimation produced by the phonetic posteriorgram (PPG) on the first two seconds of an utterance with transcript (obtained via Whisper) \emph{``finally made Yribbiade so angry that he exclaimed, those who begin the race before the signal is given up''}. (a) The raw acoustic waveform of the input speech signal. (b) The spectrogram overlaid with hard phonetic boundaries and labels, derived by taking the argmax across the $M$ classes at each frame (subject to a minimal duration constraint) to map complex acoustics onto explicit linguistic labels; shaded regions denote the corresponding macro-articulatory groups. (c) A temporal heatmap of the group-level posterior probabilities. A key strength of the PPG lies in retaining the full probability distribution rather than forcing premature classifications, capturing the continuous coarticulation and transient overlaps between adjacent sounds that characterize natural speech, visible here as the smooth probability transitions between adjacent groups.}
    \label{fig:ppg_1}
\end{figure}

Whereas phonetics provides the theoretical framework under which our explanatory model operates, the practical implementation needs to estimate the presence of phones in the acoustic speech signal. Unlike standard acoustic features which can be computed deterministically from the speech signal, phone classes cannot be directly observed. In the language of machine learning, underlying phonetic segments (phones) are simply \emph{discrete latent variables}, which can not be directly observed, but whose posterior distribution can nonetheless be estimated from the signal using a model~\cite{murphy2012machine}. In practice, we use the so-called \emph{phonetic posteriorgram} (PPG) \cite{hazen2009query}, a time-series representation where each temporal frame describes the probability of various phonetic classes being articulated at that specific moment. Conceptually, a PPG is a time-series representation where each temporal frame describes the probability of various phonetic classes being articulated at that specific moment. Unlike raw acoustic features which entangle linguistic content with speaker identity, environmental noise, channel, and various other factors, PPGs are explicitly trained to focus on the underlying phonemic content, as illustrated in Figure \ref{fig:ppg_1}.

In terms of random variables, we can formalize PPG representation as follows. Given a raw speech waveform $S$, a pre-trained acoustic model (the phonetic extractor) processes the audio to produce a temporal sequence of PPG vectors spanning $T$ frames. Let $Z_t$ represent the underlying discrete latent variable i.e., the true phonetic identity at frame $t$. The PPG vector at this frame comprises $M$ probabilities, corresponding to our predefined alphabet of $M=61$ distinct TIMIT phonetic classes. Each scalar component represents the frame-level posterior probability $P(Z_t = z_i \mid S)$, satisfying the constraint that the sum of all probabilities at frame $t$ equals one.

The probabilities of the broad phone classes defined in Table~\ref{tab:phoneme_groups} are also straightforward to obtain from a PPG representation. Noting that the $M=61$ TIMIT phone classes are mutually exclusive, the posterior of a group is simply the sum of its constituent phone probabilities. For example, the probability that a vowel is being produced in a given frame is computed as $P(\text{Vowels}\mid S) = \sum_{z_i \in \text{Vowels}} P(Z_t = z_i \mid S)$, where the summation runs over phones such as \texttt{aa}, \texttt{ae}, \texttt{iy}, and so on. The evolution of these group-level posteriors over time is visualized in the temporal heatmap of Figure~\ref{fig:ppg_1}(c). These continuous posterior estimates serve as \emph{phonetic anchors}, providing the confidence weights that guide our downstream architecture to aggregate acoustic evidence proportionally to the articulatory composition of each utterance.

\section{Probabilistic Basis of Phonetically-Interpretable Speech Deepfake Detection}
\label{section_4}

To transition deepfake speech detection from an opaque classification task into a phonetically interpretable one, we formulate the problem through a probabilistic lens. The central idea is to model the spoofing posterior $P(Y \mid X, W)$, i.e., the probability that an utterance is spoofed, conditioned jointly on a self-supervised acoustic representation $X$ and a phonetic posteriorgram $W$, and to decompose it into a sum of per-phone contributions. This decomposition does not arise mechanically; it requires a small number of explicit probabilistic assumptions about how the acoustic and phonetic streams relate to the latent phonetic identity $Z$ and to the spoofing label $Y$. By stating these assumptions upfront---concerning the sufficiency of the phonetic estimate, the acoustic dominance over phonetic-estimator errors, and the uninformative nature of phonetic priors---we obtain a decomposable classification objective whose per-phone weights and per-phone evidence terms are both architecturally addressable. The remainder of this section makes each of these assumptions and their consequences explicit, culminating in the central equation that motivates the framework's architecture.

\subsection{Definitions and Variable Formulations}

\begin{table}[h]
\centering
\caption{Summary of Random Variables and Model Parameters}
\label{tab:variables}
\begin{tabular}{lp{0.8\linewidth}}
\hline
\textbf{Variable} & \textbf{Description} \\ \hline
$S$ & The raw time-domain speech waveform. \\
$\Theta_X, \Theta_W$ & Parameter vectors (network weights) of the pre-trained acoustic and phonetic posteriorgram extractor models. \\
$X$ & Sequence-level acoustic representation, $X=\mathcal{F}_{X}(S;\Theta_{X})\in\mathbb{R}^{T_{x}\times D_{x}}$, where $T_x$ is the sequence length and $D_x$ is the feature dimension. \\
$Z$ & Latent categorical phonetic state, $Z \in \{z_1, \dots, z_M\}$. \\
$W$ & Sequence-level phone posteriorgram, $W = \mathcal{F}_W(S; \Theta_W) \in [0, 1]^{T_p \times M}$, where $T_p$ is the sequence length and $M$ represents the number of distinct phonetic classes. Each temporal row constitutes a valid frame-level categorical distribution. \\
$w_i$ & A scalar component of the phone posteriorgram at a specific temporal frame, estimating the probability of the $i$-th phonetic class: $w_i = P(Z=z_i \mid S)$. \\
$Y$ & Binary detection task label ($Y=0$: spoofed, $Y=1$: bonafide). \\ \hline
\end{tabular}
\end{table}

First, we define the observed and latent random variables for a given speech segment, as summarized in Table~\ref{tab:variables}. To explicitly contrast the two phonetic variables: both $Z$ and $W$ represent discrete categorical distributions over the $M$ possible phonetic states. For a given frame, the true latent state $Z$ represents the ideal phonetic target, functioning as a one-hot distribution concentrated on a single outcome. Conversely, $W$ can be viewed as an uncertain version of $Z$. Its soft probability estimates reflect real-world constraints present in the phone posteriorgram extractor such as modeling misspecification, finite-sample size effects, and observational noise. By distributing probability mass across multiple classes, $W$ expresses this predictive uncertainty, effectively capturing natural coarticulation without forcing a premature hard classification.

Whereas modern speech synthesis models, from early autoregressive vocoders~\cite{van2016wavenet} through mel-spectrogram--conditioned pipelines~\cite{gibiansky2017deep,shen2018natural} to contemporary adversarial, diffusion-based, and end-to-end architectures~\cite{kong2020hifi,kong2020diffwave,wang2023neural}, all aim to approximate the conditional distribution of acoustic features given linguistic targets, the task of a deepfake countermeasure is the opposite: to evaluate audio authenticity from the realized signal. As is typical in probabilistic modeling \cite{bishop2013model}, we begin with the full joint distribution over all observed and latent variables associated with a speech segment,
\begin{equation}
P(S, X, W, Z, Y),
\label{eq:joint}
\end{equation}
and ask which conditional of this joint is most relevant for the task of speech deepfake detection. Because the raw waveform $S$ enters the system only through the two deterministic feature extractors $F_X$ and $F_W$, the only quantities directly available at the decision time are the acoustic representation $X$ and the phone posteriorgram $W$. The natural object of interest is therefore the posterior of the primary class label given these two observed feature streams, namely,
\begin{equation}
P(Y \mid X, W),
\label{eq:posterior}
\end{equation}
which expresses the probability that an utterance is spoofed (equivalently, the probability that it is bonafide), conditioned on \emph{both} its acoustic content ($X$) and its phonetic-class evidence ($W$)\footnotemark[1].

\footnotetext[1]{The probability that an utterance is spoofed is denoted $P(Y = 0 \mid X, W)$, and because the two classes (bonafide and spoofed) are mutually exclusive and exhaustive, the probability that an utterance is bonafide is $P(Y = 1 \mid X, W) = 1 - P(Y = 0 \mid X, W)$.}

Most anti-spoofing systems estimate $P(Y \mid X)$ directly, relying only on the acoustic stream. The formulation here explicitly augments this with the phonetic-class evidence $W$ as a second conditioning variable, making the phonetic structure of the utterance a first-class input to the detection decision rather than something the network must implicitly recover from $X$. Operationally, $P(Y \mid X, W)$ can be parameterized by any function approximator that takes the two feature streams as input and produces a probability over the two classes---in our case, a neural architecture including a phonetic cross-attention mechanism, detailed in Section~\ref{arch}.

\subsection{Probabilistic Factorization and Assumptions}
\label{factorization_and_assumptions}

Although Eq.~\eqref{eq:posterior} fully specifies the detection problem, it offers little insight into \emph{why} a given decision is reached. To expose the role of phonetic structure in this decision, we reintroduce the latent variable $Z$ that represents the true categorical phonetic state underlying the frame, and marginalize over it. Because $Z$ takes one of $M$ discrete values, the law of total probability \cite{murphy2012machine} together with the chain rule yields an exact, unapproximated factorization:
\begin{align}
P(Y \mid X, W)
&= \sum_{i=1}^{M} P(Y, Z=z_i \mid X, W) \nonumber \\
&= \sum_{i=1}^{M} P(Y \mid X, W, Z=z_i)
\cdot P(Z=z_i \mid X, W).
\label{eq:full_expansion}
\end{align}

Importantly, this factorization decomposes the detection posterior into two interpretable components: a \emph{phone-conditional spoofing term} $P(Y \mid X, W, Z=z_i)$, which scores how authentic the acoustic evidence appears given a hypothesized phonetic state; and a \emph{phonetic responsibility term} $P(Z=z_i \mid X, W)$, which assigns probability to each phonetic state. To render Eq.~\eqref{eq:full_expansion} tractable for our neural architecture, we assert three probabilistic assumptions based on the physical mechanics of speech synthesis and feature extraction, summarized as follows.

\begin{tcolorbox}[colback=blue!5!white,colframe=blue!75!black,title=Modeling Assumptions for Probabilistic Factorization]

\begin{itemize}
    \item \textbf{Assumption 1 (A1): Sufficiency of the Phonetic Estimate.} We assume the term $P(Z=z_i \mid X, W)$ simplifies to $w_i$. Because the phonetic estimator $W$ is explicitly optimized to predict the true phoneme directly from the raw signal ($w_i = P(Z=z_i \mid S)$), conditioning additionally on the abstracted acoustic feature $X$ is assumed to provide negligible added phonetic information.
    
    \item \textbf{Assumption 2 (A2): Acoustic Sufficiency Over Phonetic-Estimator Errors.} Our second assumption asserts that $P(Y \mid X, W, Z=z_i) = P(Y \mid X, Z=z_i)$. In modern speech synthesis research, objective and subjective intelligibility evaluations reveal that generative models often fail to perfectly reproduce intended phonetic targets \cite{cooper2024review, perrotin2025refining}. Therefore, the divergence between the target $Z$ and the estimated $W$ constitutes a theoretical spoofing artifact. However, we assume that the rich, high-dimensional acoustic representation $X$ predominantly captures the low-level spectral and temporal distortions. Thus, given $X$ and $Z$, we assert that $W$ provides marginal added information about the class label $Y$.
    
    \item \textbf{Assumption 3 (A3): Uninformative Phonetic Priors.} We assume $P(Y \mid Z=z_i) = P(Y)$, i.e., $Y$ and $Z$ are statistically independent.. Because deepfake systems are explicitly trained to mimic human phonemic distributions, knowing that a specific phoneme is present in an utterance provides no inherent evidence regarding the audio's authenticity. An empirical justification of this assumption is given in~\ref{app:assumption3}.
\end{itemize}

\end{tcolorbox}

\noindent
Applying Assumptions~1 and 2 to Eq.~\eqref{eq:full_expansion} reduces the exact factorization to a tractable form, as follows:

\begin{align}
P(Y \mid X, W)
&= \sum_{i=1}^{M} P(Y \mid X, W, Z=z_i)\,\cdot\, P(Z=z_i \mid X, W)
&& \text{[By Eq.~\eqref{eq:full_expansion}]} \\
&= \sum_{i=1}^{M} P(Y \mid X, W, Z=z_i)\,\cdot\, w_i
&& \text{(by A1)} \\
&= \sum_{i=1}^{M} P(Y \mid X, Z=z_i)\,\cdot\, w_i
&& \text{(by A2)}
\label{eq:derivation_final}
\end{align}

\noindent
This yields the central probabilistic decomposition that our architecture is designed to operationalize:

\begin{equation}
\boxed{
P(Y \mid X, W)
=
\sum_{i=1}^{M}
\underbrace{w_i}_{\text{phonetic presence weight}}
\cdot
\underbrace{P(Y \mid X, Z = z_i)}_{\text{spoofing score}}
}
\label{eq:final_factorization}
\end{equation}

\begin{figure*}[h!]
\centering
\begin{tikzpicture}[
    >=Stealth,
    font=\sffamily\small,
    box/.style={rectangle, draw=gray!60, thick, rounded corners, align=center, minimum height=0.8cm, fill=gray!5},
    evidence_box/.style={rectangle, draw=blue!70, thick, rounded corners, align=center, fill=blue!5, minimum width=2.1cm, minimum height=1cm, font=\normalsize},
    weight_box/.style={rectangle, draw=orange!80, thick, rounded corners, align=center, fill=orange!10, minimum width=1.2cm, minimum height=1cm, font=\normalsize},
    sum_node/.style={circle, draw=red!80, thick, fill=red!10, minimum size=1cm, align=center},
    mul_node/.style={circle, draw=gray!80, thick, fill=gray!10, minimum size=0.6cm, align=center, inner sep=0pt},
    arrow/.style={->, thick, draw=gray!70},
    blue_arrow/.style={->, thick, draw=blue!60},
    orange_arrow/.style={->, thick, draw=orange!80},
    blue_line/.style={thick, draw=blue!60},
    orange_line/.style={thick, draw=orange!80}
]

\node[box, fill=gray!15] (audio) at (0, 0) {\textbf{Raw Audio Waveform} ($S$)};
\node[box] (xlsr) at (-2.8, -1.5) {\textbf{Acoustic Features}\\($X$)};
\node[box] (ppg) at (2.8, -1.5) {\textbf{Phonetic Probabilities}\\($W$)};

\draw[arrow] (audio.south) -- ++(0,-0.25) -| (xlsr.north);
\draw[arrow] (audio.south) -- ++(0,-0.25) -| (ppg.north);

\draw[blue_line] (xlsr.south) -- ++(0, -0.6) coordinate (Xbus);
\draw[orange_line] (ppg.south) -- ++(0, -1.0) coordinate (Wbus);

\def\gA{-5.5}
\def\gB{-1.85}
\def\gC{1.85}
\def\gD{5.5}
\def\yBox{-3.8} 
\def\yMul{-5.2} 
\def\boxSep{0.88}

\node[evidence_box] (e_vowel) at (\gA - \boxSep, \yBox) {$P(Y|X, z_{\text{vwl}})$\\{\small Evidence: 0.5}};
\node[weight_box] (w_vowel) at (\gA + \boxSep, \yBox) {$w_{\text{vwl}}$\\{\small 0.40}};
\node[mul_node] (mul1) at (\gA, \yMul) {$\times$};

\node[evidence_box] (e_stop) at (\gB - \boxSep, \yBox) {$P(Y|X, z_{\text{stp}})$\\{\small Evidence: 0.9}};
\node[weight_box] (w_stop) at (\gB + \boxSep, \yBox) {$w_{\text{stp}}$\\{\small 0.15}};
\node[mul_node] (mul2) at (\gB, \yMul) {$\times$};

\node[evidence_box] (e_fric) at (\gC - \boxSep, \yBox) {$P(Y|X, z_{\text{frc}})$\\{\small Evidence: 0.8}};
\node[weight_box] (w_fric) at (\gC + \boxSep, \yBox) {$w_{\text{frc}}$\\{\small 0.20}};
\node[mul_node] (mul3) at (\gC, \yMul) {$\times$};

\node[evidence_box] (e_other) at (\gD - \boxSep, \yBox) {$P(Y|X, z_{\text{oth}})$\\{\small Evidence: 0.2}};
\node[weight_box] (w_other) at (\gD + \boxSep, \yBox) {$w_{\text{oth}}$\\{\small 0.25}};
\node[mul_node] (mul4) at (\gD, \yMul) {$\times$};

\draw[blue_arrow, rounded corners=3pt] (Xbus) -| (e_vowel.north);
\draw[blue_arrow, rounded corners=3pt] (Xbus) -| (e_stop.north);
\draw[blue_arrow, rounded corners=3pt] (Xbus) -| (e_fric.north);
\draw[blue_arrow, rounded corners=3pt] (Xbus) -| (e_other.north);

\draw[orange_arrow, rounded corners=3pt] (Wbus) -| (w_vowel.north);
\draw[orange_arrow, rounded corners=3pt] (Wbus) -| (w_stop.north);
\draw[orange_arrow, rounded corners=3pt] (Wbus) -| (w_fric.north);
\draw[orange_arrow, rounded corners=3pt] (Wbus) -| (w_other.north);

\draw[blue_arrow] (e_vowel.south) -- (mul1.north west);
\draw[orange_arrow] (w_vowel.south) -- (mul1.north east);

\draw[blue_arrow] (e_stop.south) -- (mul2.north west);
\draw[orange_arrow] (w_stop.south) -- (mul2.north east);

\draw[blue_arrow] (e_fric.south) -- (mul3.north west);
\draw[orange_arrow] (w_fric.south) -- (mul3.north east);

\draw[blue_arrow] (e_other.south) -- (mul4.north west);
\draw[orange_arrow] (w_other.south) -- (mul4.north east);

\node[sum_node] (sum) at (0, -6.5) {$\boldsymbol{\Sigma}$};

\draw[arrow, rounded corners=5pt] (mul1.south) |- (sum.west);
\draw[arrow] (mul2.south) -- (sum.north west);
\draw[arrow] (mul3.south) -- (sum.north east);
\draw[arrow, rounded corners=5pt] (mul4.south) |- (sum.east);

\node[box, fill=green!10, draw=green!80, thick, minimum height=1cm] (output) at (0, -8) {\textbf{Final Spoofing Score}\\$P(Y|X,W)$};
\draw[arrow] (sum.south) -- (output.north);

\end{tikzpicture}
\caption{The proposed phonetically transparent scoring framework. The utterance's final spoofing score is decomposed into phoneme-specific acoustic evidence (blue boxes) modulated by the continuous phonetic presence weights (orange boxes).}
\label{fig:dreamlike_framework}
\end{figure*}
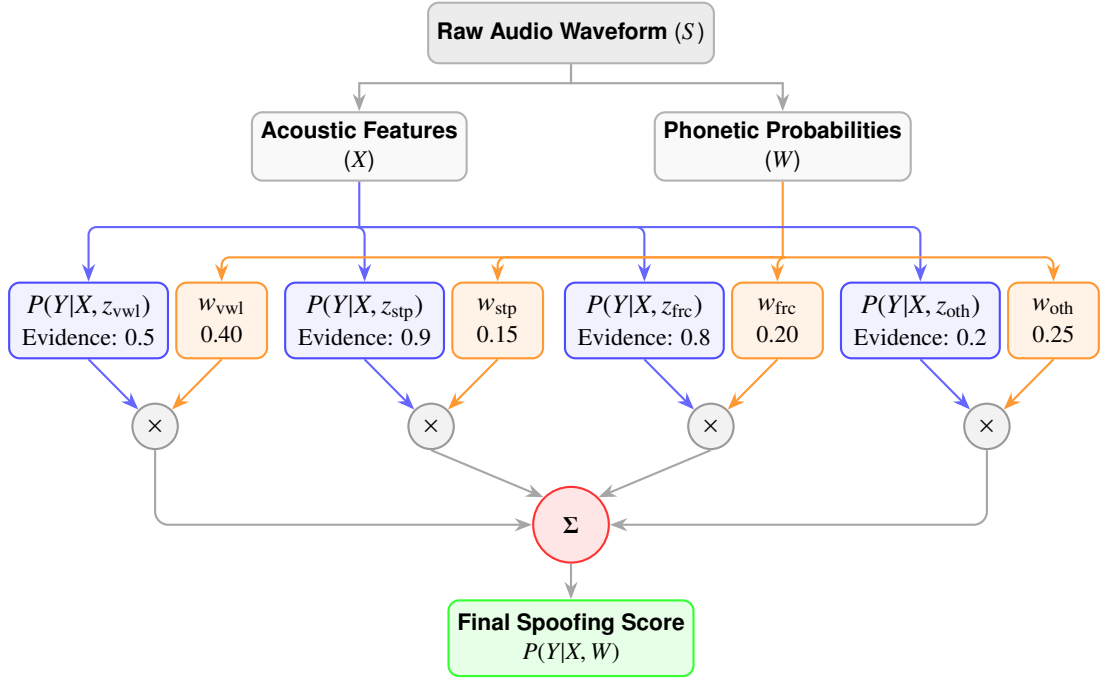

Equation~\eqref{eq:final_factorization} demonstrates that the total spoofing score can be decomposed into a \textbf{phonetically decomposable detection score}: a weighted summation of independent, phone-specific spoofing scores $P(Y \mid X, Z = z_i)$ modulated by the continuous phonetic presence weights $w_i$ (see Figure~\ref{fig:dreamlike_framework}). It is important to note that both the weights $w_i$ and the per-phone scores $P(Y \mid X, Z = z_i)$ are \emph{utterance-specific}: at every frame of the utterance, they are computed from the local acoustic content $X$ and the local phonetic posterior $W$. Therefore, neither component is a fixed property of a phone class but is instead determined dynamically by each input utterance.

\textbf{Example.} To demonstrate use of ~\eqref{eq:final_factorization}, consider a hypothetical, simplified scenario in which our detection model analyzes a single speech frame using an alphabet of only three phones ($M=3$): a vowel (\texttt{a}), a plosive (\texttt{p}), and silence (\texttt{pau}). If the phonetic estimator is 70\% confident the frame contains the vowel, 20\% confident it contains the plosive, and 10\% confident it is silence, the overall authenticity probability expands to:
\begin{equation}
P(Y \mid X, W)
=
0.70 \cdot P(Y \mid X, Z=\texttt{a})
+
0.20 \cdot P(Y \mid X, Z=\texttt{p})
+
0.10 \cdot P(Y \mid X, Z=\texttt{pau}).
\label{eq:toy_expansion}
\end{equation}

Each term $P(Y \mid X, Z=z_i)$ here is interpreted as the probability that the observed acoustic frame is authentic \emph{conditional on it being a realization of phone $z_i$}: $P(Y \mid X, Z=\texttt{a})$ scores whether $X$ looks like a genuinely produced vowel, $P(Y \mid X, Z=\texttt{p})$ scores whether it looks like a genuinely produced plosive, and so on. The final detection score is therefore not a single opaque number but a weighted combination of three phone-conditional authenticity judgments.

This is the essential contrast with conventional anti-spoofing detectors, which produce a single utterance-level score $P(Y \mid X)$ with no explicit awareness of which speech sounds are present at any given moment. Such phone-blind scoring conflates evidence from all phonetic content into one decision, making it impossible to determine \emph{where} or \emph{why} the model finds an utterance suspicious. The \emph{phonetically decomposed score}, in Equation~\eqref{eq:final_factorization}, by contrast, makes the contribution of each phonetic state explicit: the final decision is proportionally driven by the acoustic evidence for the most probable speech sounds, the per-phone scores can be inspected individually to identify which sounds carry the discriminative signal, and the model explicitly accounts for natural articulatory uncertainty rather than committing to a single hard phonetic interpretation.

\subsection{Discriminative Power vs. Occurrence Frequency}

Having established the phonetically decomposed score, it is instructive to disentangle a phone's natural occurrence frequency from its discriminative utility. We aim to demonstrate formally that simply because a specific speech sound appears frequently in an utterance, it does not inherently provide strong evidence of synthesis artifacts. Equivalently, a detector that fails to separate phonetic \emph{prevalence} from phonetic \emph{informativeness} risks aligning its decisions with the marginal phone distribution rather than with the acoustic evidence of spoofing.

To make this distinction explicit, we apply Bayes' theorem to the phone-conditional spoofing score from Eq.~\eqref{eq:final_factorization}, expanding the joint $P(X, Z=z_i \mid Y)$ via the product rule and similarly factorizing the normalizer $P(X, Z=z_i)$:

\begin{align}
P(Y \mid X, Z = z_i)
&= \frac{P(X, Z = z_i \mid Y)\cdot P(Y)}{P(X, Z = z_i)}
\label{eq:bayes_step1}
\\[4pt]
&= \frac{P(X \mid Y, Z = z_i)\cdot P(Z = z_i \mid Y)\cdot P(Y)}
{P(X \mid Z = z_i)\cdot P(Z = z_i)}
\label{eq:bayes_step2}
\\[4pt]
&= \frac{P(X \mid Y, Z = z_i)\cdot P(Y \mid Z = z_i)}
{P(X \mid Z = z_i)}.
\label{eq:bayes_exact}
\end{align}

By directly applying Assumption 3 to Equation \eqref{eq:bayes_exact}, we substitute the phonetic prior $P(Y \mid Z=z_i)$ with the global prior $P(Y)$. Because this global prior acts as a constant scaling factor across all phonetic classes, independent of the acoustic and phonetic variables, we can drop it to express the discriminative relationship as a direct proportionality:

\begin{equation}P(Y \mid X, Z=z_i) \propto \frac{P(X \mid Y, Z=z_i)}{P(X \mid Z=z_i)} \label{eq:bayes_proportional}\end{equation}

Equation \eqref{eq:bayes_exact} demonstrates that the discriminative power stems entirely from the class-conditional acoustic likelihood, $P(X \mid Y, Z=z_i)$. Because the denominator $P(X \mid Z=z_i)$ represents the marginal acoustic distribution of the phoneme independent of its class label $Y$, it functions merely as a normalizing constant with respect to the classification decision. Therefore, the ability to distinguish between bonafide and spoofed speech relies exclusively on the numerator. This term mathematically quantifies the divergence between the acoustic distribution of a human uttering a specific phoneme (e.g., $P(X \mid Y=1, Z=/\text{a}/)$) versus a generative model synthesizing that same phoneme (e.g., $P(X \mid Y=0, Z=/\text{a}/)$)

Under the probabilistic assumptions asserted above, this mathematical distinction proves that a phoneme's high natural occurrence frequency does not guarantee high discriminative utility. Consider another hypothetical scenario where a generative model perfectly replicates a highly frequent phoneme. In this case, the bonafide and spoofed class-conditional acoustic distributions overlap entirely, i.e. $P(X \mid Y=1, Z=z_i) = P(X \mid Y=0, Z=z_i)$. Consequently, the local acoustic evidence at this phone yields a posterior that simply equals the prior ($P(Y \mid X, Z = z_i) = P(Y)$), and the phone therefore provides no additional class-discriminative information beyond the prior, despite its frequent occurrence throughout the utterance. Conversely, if a generative model systematically fails to replicate the complex dynamics of a rare phoneme such as the persistent spectral misrendering of obstruents and nasals observed in modern end-to-end text-to-speech architectures \cite{fullwood2026mind}, the resulting distributions will be highly distinct, yielding powerful spoofing cues.

Consequently, this principled separation exposes a fundamental flaw in any aggregation mechanism that treats temporal frames as interchangeable units. The simplest such mechanism, \emph{mean temporal pooling}, collapses a sequence of frame-level features or scores into an utterance-level decision by uniform averaging across time. Although more elaborate alternatives such as moment-based pooling~\cite{snyder2018x}, self-attentive pooling~\cite{okabe2018attentive}, and graph-attention aggregation~\cite{jung2022aasist} learn input-dependent weights, all of them remain agnostic to the phonemic content of the utterance. Their weights are determined by acoustic salience or learned attention scores rather than by the phonetic identity of each frame. As a result, frequent but phonetically uninformative segments can still dominate the aggregated score. To recover the per-phone weighting that Eq.~\eqref{eq:bayes_proportional} identifies as the optimal classification objective, we therefore design our aggregation mechanism to decouple occurrence frequency from discriminative power, dynamically weighting the acoustic evidence based on the phone-conditional divergence between bonafide and spoof distributions.

\section{Proposed Phoneme-Guided Cross-Attention Architecture}
\label{arch}

\begin{figure}[h!]
    \centering
    \includegraphics[width=\linewidth]{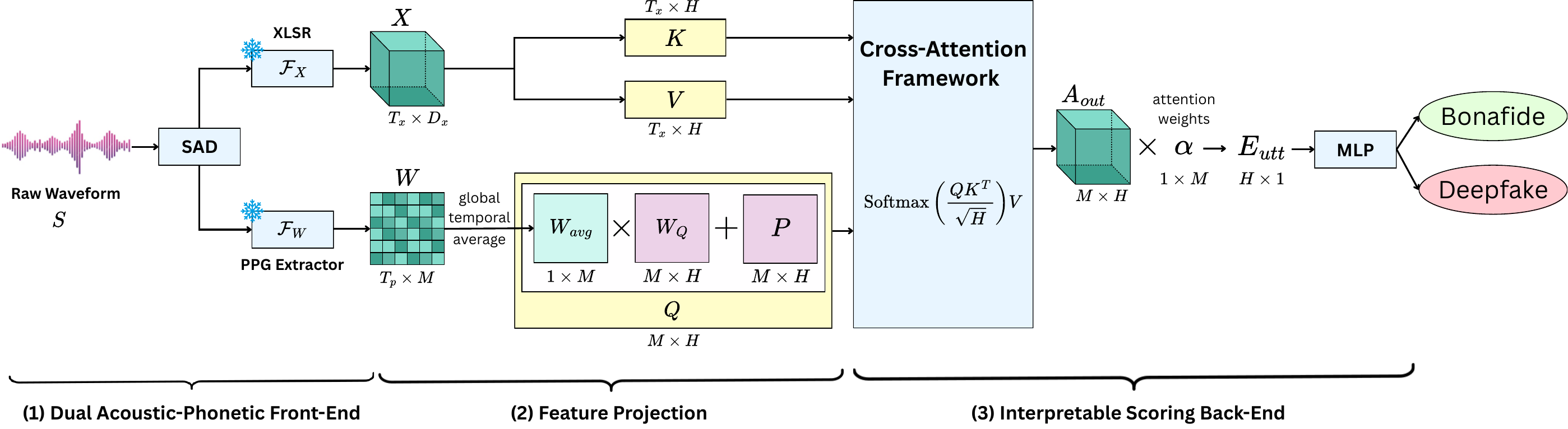}
    \caption{Architecture of the end-to-end phoneme-guided cross-attention framework. The system processes raw audio ($S$) through three distinct phases to yield an interpretable authenticity score. In the \textbf{(1) Dual Acoustic-Phonetic Front-End}, speech-activated audio is processed by an XLSR model to extract continuous acoustic Keys ($K$) and Values ($V$), while a parallel PPG extractor yields sequence-level phonetic probabilities ($W$) that are temporally averaged and combined with learnable prototypes ($P$) to generate discrete phonetic Queries ($Q$). In the \textbf{(2) Cross-Attention Framework}, these queries probe the acoustic features via scaled dot-product attention, yielding a phoneme-aligned acoustic representation matrix ($A_{\text{out}}$). Finally, in the \textbf{(3) Interpretable Scoring Back-End}, $A_{\text{out}}$ is projected via a pooling vector to compute raw phoneme logits. Then, a softmax operation converts these logits into interpretable structural weights ($\alpha$). These weights guide a weighted summation of $A_{\text{out}}$ to form a global utterance embedding ($E_{\text{utt}}$), which is then processed by a classification MLP to output the final binary prediction.}
    \label{fig:architecture_diagram}
\end{figure}

Equipped with the theoretical probabilistic factorization (Section~\ref{section_4})
and the continuous phonetic estimators provided by the PPGs (Section~\ref{section_3_2}),
we now detail our phoneme-guided cross-attention neural architecture that operationalizes the
decomposable spoofing score of Eq.~\eqref{eq:final_factorization}. As illustrated in
Figure~\ref{fig:architecture_diagram}, the architecture consists of three components: (1) a dual acoustic--phonetic front-end, (2) a cross-attention framework that
realizes the phone-conditional spoofing score $P(Y \mid X, Z = z_i)$, and (3) an
interpretable scoring back-end that realizes the phone-presence weights $w_i$ and
produces the final utterance-level decision.

\begin{figure}[h!]
    \centering
    \includegraphics[width=\linewidth]{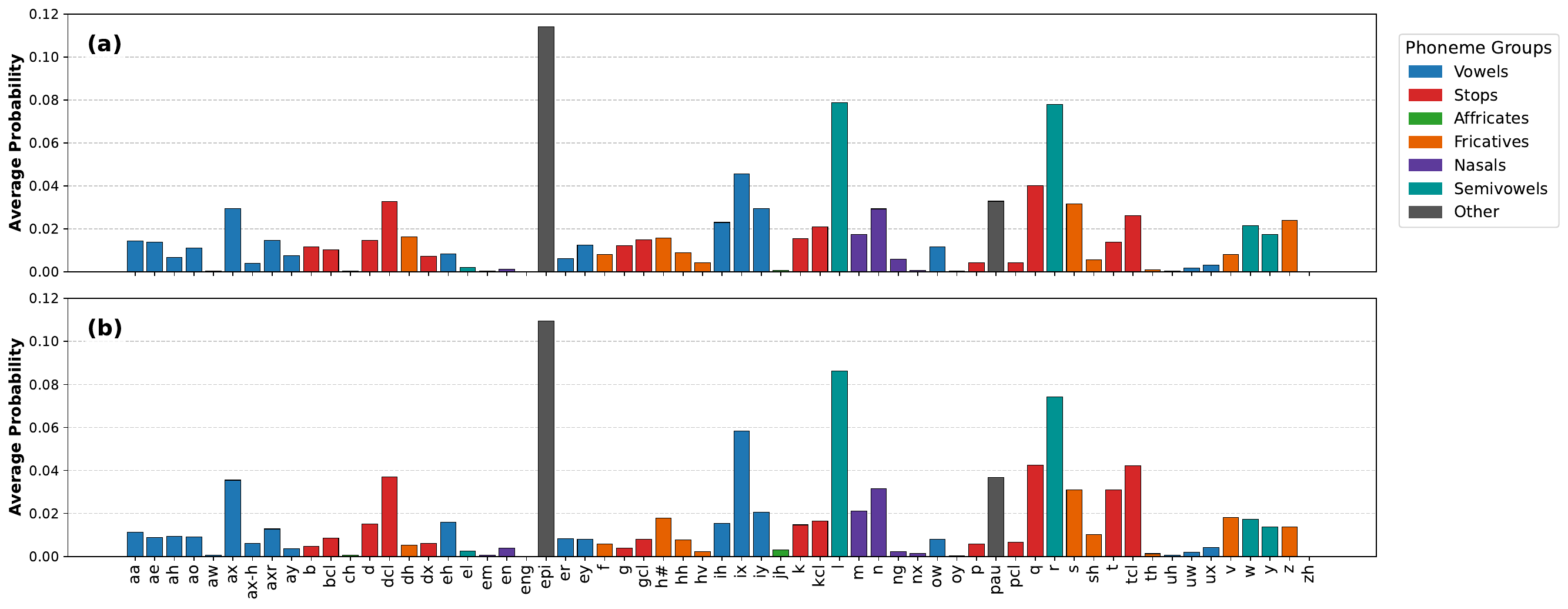}
    \caption{Visualization of the utterance-level phonetic-context vector $W_{\text{avg}} \in \mathbb{R}^{1 \times M}$ for two example utterances from the same speaker, drawn from the ASVspoof~5 corpus. Here, $W_{\text{avg}}$ denotes the global temporal average of the PPG sequence, summarizing the overall phonetic composition of an utterance as a single probability distribution over the 61 TIMIT phone classes. (a) corresponds to the utterance \emph{``finally made Yribbiade so angry that he exclaimed, those who begin the race before the signal is given up.''} (b) \emph{corresponds to the utterance ``That's all the vans who made no movement to suggest that she had observed it.''}}
    \label{fig:W_avg}
\end{figure}

\subsection{Dual Acoustic--Phonetic Front-End}
\label{subsec:frontend}

Block~\textbf{(1)} of Figure~\ref{fig:architecture_diagram} extracts the acoustic and the phonetic features defined in Section~\ref{section_4}. The raw waveform $S$ is first passed through a
speech-activity detector (SAD) and then through two frozen, parallel extractors: a
self-supervised acoustic model $\mathcal{F}_X$ (XLS-R) and a PPG extrator $\mathcal{F}_W$. These yield, respectively, a sequence of acoustic embeddings
$X \in \mathbb{R}^{T_x \times D_x}$ and a phonetic posteriorgram
$W \in \mathbb{R}^{T_p \times M}$. Here, $T_x$ and $T_p$ are the temporal sequence
lengths and $M = 61$ is the size of the phonetic alphabet. The specific
instantiations of $\mathcal{F}_X$ and $\mathcal{F}_W$, together with all
preprocessing details, are
deferred to Section~\ref{section_experiments}.

\subsection{Cross-Attention Framework}
\label{subsec:crossattn}

Block~\textbf{(2)} of Figure~\ref{fig:architecture_diagram} bridges the acoustic and phonetic
streams through a cross-attention mechanism. We briefly recall the
underlying operation before describing our specific use case. \emph{Attention}~\cite{vaswani2017attention}
computes a weighted sum of \emph{value} vectors, with weights determined by the
similarity between \emph{query} and \emph{key} vectors. In \emph{self}-attention,
queries, keys, and values are all derived from a single input sequence, allowing
the model to capture contextual relationships within one
domain~\cite{vaswani2017attention,devlin2019bert}. In \emph{cross}-attention, by
contrast, the queries are derived from one sequence while keys and values come from
another, allowing the queries to selectively probe information in the second
sequence. 
This asymmetry is what makes cross-attention a natural fit when one structured stream must be interpreted in the context of another---speech translation~\cite{bahdanau2014neural}, sequence-to-sequence speech recognition decoders~\cite{chan2016listen}, and conditional speech synthesis~\cite{shen2018natural} all involve mapping between two distinct domains (text and audio in various directions), where one stream legitimately poses a question and the other supplies the evidence used to answer it. Cross-attention is therefore a known,
off-the-shelf mechanism studied well across different applications; the contribution of our work is not the mechanism
itself but its use as a acoustic--phonetic bridge for explainable spoofing detection.

We use the cross-attention mechanism asymmetrically: the acoustic embedding $X$ supplies the keys and values, while the phonetic information supplies the queries. Two learned linear projections of $X$
yield
\begin{equation}
K = X W_K \in \mathbb{R}^{T_x \times H},
\qquad
V = X W_V \in \mathbb{R}^{T_x \times H},
\label{eq:keys_values}
\end{equation}
where $H$ denotes the attention hidden dimension. The matrices $K$ and $V$
encapsulate the temporal acoustic content, including artifacts left by
neural vocoders or voice conversion algorithms, which the phonetic queries, explained next, will probe.

The query construction is the first component specific to our framework. A simple design would represent each of the $M$ phonetic classes by a fixed learnable
embedding common to all utterance. Such global prototype, however, ignores the speaker, prosody, and recording
conditions of the input utterance. We therefore combine a global, phone-specific component with an utterance-specific contextual component, and we refer to the resulting combined per-class query vectors as \emph{phonetic anchors}. The global
component is a learnable prototype matrix $\probP \in \mathbb{R}^{M \times H}$, containing one row per phonetic class. The utterance-specific component, in turn, is obtained by
temporally averaging the posteriorgram $W$,
\begin{equation}
W_{\mathrm{avg}}
=
\frac{1}{T_p}\sum_{t=1}^{T_p} W_{t}
\;\in\; \mathbb{R}^{1 \times M},
\label{eq:w_avg}
\end{equation}
which summarizes the phonemic content of the utterance. After projecting $W_{\mathrm{avg}}$ to the attention hidden dimension
through $W_Q \in \mathbb{R}^{M \times H}$ and broadcasting across the $M$
prototypes, the final query matrix becomes

\begin{equation}
Q = \underbrace{\probP}_{\substack{\text{global,} \\ \text{learned during training}}}
\;+\;
\underbrace{W_{\mathrm{avg}} W_Q}_{\substack{\text{utterance-specific} \\ \text{adaptive bias}}}
\;\in\; \mathbb{R}^{M \times H}.
\label{eq:queries}
\end{equation}

Each row of $Q$ is thus a phonetic anchor that is semantically tied to one
phonetic class through $P$ and contextually grounded in the present utterance
through $W_{\mathrm{avg}} W_Q$. The utterance context is injected here, in the
queries, rather than into $K$ or $V$: this preserves the interpretation of each
row of $A_{\mathrm{out}}$ (defined below) as the acoustic evidence associated
with one phonetic class.

With $Q$, $K$, and $V$ defined, the cross-attention block computes\footnotemark[2]

\begin{equation}
A_{\mathrm{out}}
=
\mathrm{Softmax}\!\left(\frac{QK^\top}{\sqrt{H}}\right) V
\;=\;
\underbrace{
\left[A_{\mathrm{out}}[1], \ldots, A_{\mathrm{out}}[M]\right]^{\top}
}_{\substack{
\text{$m$-th row: neural estimate of} \\
P(Y \mid X,~Z=z_m) \text{ from Eq.~\eqref{eq:final_factorization}}
}}
\;\in\; \mathbb{R}^{M \times H},
\label{eq:Aout}
\end{equation}
following the standard \emph{scaled dot-product}
formulation~\cite{vaswani2017attention}. Note the shape of $A_{\mathrm{out}}$: by construction it has exactly $M$ rows, one per
phonetic class, where the $m$-th row aggregates the portions of the acoustic
stream $V$ that the corresponding phonetic anchor $Q_m$ attended to. This
per-row structure is precisely how the architecture realizes the
phone-conditional spoofing score $P(Y \mid X, Z = z_i)$ of
Eq.~\eqref{eq:final_factorization}: each row $A_{\mathrm{out}}[m]$ is a neural
estimate of the phone-conditional evidence vector for class $m$, computed from
the same acoustic representation $X$ but isolated to the portion of $X$ that is
articulatorily consistent with phone $z_m$.

\footnotetext[2]{$\mathrm{Softmax}(z)_i = \frac{\exp(z_i)}{\sum_{j=1}^{M} \exp(z_j)}$
maps a vector of real-valued logits to a probability distribution that sums to one.}

\subsection{Interpretable Scoring Back-End}
\label{subsec:backend}

Finally, block~\textbf{(3)} of Figure~\ref{fig:architecture_diagram} aggregates the $M$
phone-conditional evidence rows of $A_{\mathrm{out}}$ into a single
utterance-level score. This aggregation step is where the phone-presence
weights $w_i$ of Eq.~\eqref{eq:final_factorization} are explicitly realized, and we evaluate two variants that differ in how those weights are determined: one that treats all phonetic classes as equally informative, and one in which the weights are learned end-to-end from the data.

\paragraph{\textbf{Mean pooling}}
The simplest variant averages the $M$ rows uniformly,
\begin{align}
E_{\mathrm{utt}}
&= \frac{1}{M}\sum_{m=1}^{M} A_{\mathrm{out}}[m] \nonumber\\
&= \sum_{m=1}^{M} \frac{1}{M}\, A_{\mathrm{out}}[m].
\label{eq:mean_pool}
\end{align}

Comparing Eq.~\eqref{eq:mean_pool} with Eq.~\eqref{eq:final_factorization},
mean pooling corresponds to the particular choice $w_i = 1/M$ for all
$i = 1, \dots, M$---a flat prior that assigns identical importance to every
phonetic class regardless of its presence or discriminative power in the
current utterance. This choice contradicts the central result of
Section~\ref{section_4}: frequent but uninformative phones should not be
allowed to dominate. Likewise, rare but highly discriminative phones should not be
down-weighted. We therefore retain mean pooling merely for reference purposes.

It is worth distinguishing the uniform weighting in Eq.~\eqref{eq:mean_pool}
from the input-dependent attention weights inside the cross-attention block of
Eq.~\eqref{eq:Aout}. These attention weights determine \emph{how the acoustic
content of $X$ is distributed across the phonetic anchors} (i.e., how each
$A_{\mathrm{out}}[m]$ is constructed), whereas the pooling weights determine
\emph{how the resulting phone-conditional evidence rows are combined into an
utterance-level decision}. Replacing uniform pooling weights by data-driven
ones is therefore complementary to, not subsumed by, the cross-attention step.

\paragraph{\textbf{Learned weighted pooling}}
To operationalize the data-driven $w_i$ implied by
Eq.~\eqref{eq:final_factorization}, we replace the uniform weights of
Eq.~\eqref{eq:mean_pool} with weights computed from $A_{\mathrm{out}}$ itself.
A learned linear scoring layer $W_p \in \mathbb{R}^{H \times 1}$ maps each row
to a scalar logit,
\begin{equation}
L_p = A_{\mathrm{out}} W_p
\;\in\; \mathbb{R}^{M},
\label{eq:pool_logits}
\end{equation}
and a softmax over $L_p$ yields normalized phone-presence weights
\begin{equation}
\alpha = \mathrm{Softmax}(L_p)
\;\in\; \mathbb{R}^{M}.
\label{eq:alpha}
\end{equation}
\noindent
The utterance embedding is then
\begin{equation}
E_{\mathrm{utt}}
=
\sum_{m=1}^{M}\alpha_m\, A_{\mathrm{out}}[m],
\label{eq:weighted_pool}
\end{equation}
which is the direct architectural counterpart of
Eq.~\eqref{eq:final_factorization}: $\alpha_m$ plays the role of $w_i$, and
$A_{\mathrm{out}}[m]$ plays the role of $P(Y \mid X, Z = z_i)$.

\paragraph{\textbf{Classification head and training}}
$E_{\mathrm{utt}}$ (from either pooling variant) is passed through a multi-layer
perceptron followed by a sigmoid, producing the probability that the
utterance is spoofed. Apart from the frozen XLSR and PPG extractors, the full system is trained with binary
cross-entropy loss; under the weighted-pooling variant, $\alpha$ is therefore
learned implicitly through the classification gradient and converges to assign
mass to the phonetic classes that consistently provide the strongest separation
between bonafide and spoofed speech. The complete end-to-end training procedure is summarized in
Algorithm~\ref{alg:cross_attention}. To ensure reproducibility, full implementation details and source code are publicly available\footnotemark[3].

\footnotetext[3]{\url{https://github.com/Manasi2001/Phonetically-Explainable-Speech-Deepfake-Detection}}

\paragraph{\textbf{Targeted phonetic masking}}
A central interpretive question that the proposed framework is uniquely positioned to address is which articulatory groups actually carry the bulk of the discriminative signal---that is, whether all phonetic categories contribute equally to a detection decision, or whether a small subset of them is responsible for most of the model's discriminative power. To address this question, we exploit the fact that $\alpha$ is computed through an explicit softmax over $M$ phone-aligned logits, which allows us to selectively suppress the contribution of any chosen subset of phonetic classes at inference time. Given
any subset $\mathcal{M} \subseteq \{1,\dots,M\}$ of phonetic classes, we can
set $L_p[m] \leftarrow -\infty$ for all $m \in \mathcal{M}$ before applying the
softmax. The resulting $\alpha$ has zero mass on the masked classes, which effectively forces the detector to ignore the corresponding phonetic-class evidence. This
will be used in Section~\ref{section_experiments} to empirically isolate which
phonetic groups carry the dominant discriminative signal.\newline

\begin{algorithm}[H]
\caption{Phoneme-Guided Cross-Attention for Speech Deepfake Detection}
\label{alg:cross_attention}

\SetAlgoLined
\KwIn{Raw speech dataset $\mathcal{D} = \{(S_n, y_n)\}_{n=1}^N$ with binary labels $y_n \in \{0, 1\}$, Pre-trained PPG extractor model $\mathcal{F}_{W}$, Pre-trained acoustic model $\mathcal{F}_{X}$, Set of target phoneme indices to mask $\mathcal{M}$ (empty for full model).}
\KwOut{Trained parameters $\Theta$.}

\BlankLine
\tcp{Phase 1: Feature Extraction (Pre-processing)}
\For{each $(S_n, y_n) \in \mathcal{D}$}{
    $S_n' \leftarrow \text{SAD\_TrimSilence}(S_n)$ \tcp*{Remove non-speech frames}
    $X_n \leftarrow \mathcal{F}_{X}(S_n') \in \mathbb{R}^{T_x \times D_x}$ \tcp*{Extract XLSR embeddings}
    $W_n \leftarrow \text{Softmax}(\mathcal{F}_{W}(S_n')) \in \mathbb{R}^{T_p \times M}$ \tcp*{Extract PPGs ($M=61$ phonemes)}
}

\BlankLine
\tcp{Phase 2: Model Initialization}
Initialize learnable prototypes $\probP \in \mathbb{R}^{M \times H}$\;
Initialize projection matrices $W_K, W_V \in \mathbb{R}^{D_x \times H}$, $W_Q \in \mathbb{R}^{M \times H}$\;
Initialize pooling vector $W_p \in \mathbb{R}^{H \times 1}$ and Classifier MLP layers\;
Let $\Theta$ be the set of all learnable parameters\;

\BlankLine
\tcp{Phase 3: Training Loop}
\For{$\text{epoch} \leftarrow 1$ \KwTo $\text{MaxEpochs}$}{
    \For{each batch $(X, W, y) \in \mathcal{D}$}{
        \tcp{Step 3a: Cross-Attention Formulation}
        $K \leftarrow X W_K$, \quad $V \leftarrow X W_V$ \tcp*{Acoustic Keys and Values}
        $W_{avg} \leftarrow \frac{1}{T_p} \sum_{t=1}^{T_p} W_{t}$ \tcp*{Global utterance phonetic context}
        $Q \leftarrow \probP + (W_{avg} W_Q)$ \tcp*{Queries: Static Prototypes + Context}
        
        \BlankLine
        \tcp{Compute Multi-Head Attention (MHA)}
        $A_{out} \leftarrow \text{Softmax}\left(\frac{Q K^T}{\sqrt{H}}\right) V \in \mathbb{R}^{M \times H}$\;
        
        \BlankLine
        \tcp{Step 3b: Pooling}
        $L_p \leftarrow A_{out} W_p \in \mathbb{R}^{M}$ \tcp*{Compute raw pooling logits}
        
        
        $\alpha \leftarrow \text{Softmax}(L_p)$ \tcp*{Calculate interpretable phonetic weights}
        $E_{utt} \leftarrow \sum_{m=1}^{M} \alpha_m \cdot A_{out}[m]$ \tcp*{Utterance embedding (Weighted Sum)}
        
        \BlankLine
        \tcp{Step 3c: Classification and Optimization}
        $\hat{y} \leftarrow \text{Classifier}(E_{utt})$ \tcp*{Forward pass through MLP}
        $\mathcal{L} \leftarrow - \left[ y \log(\sigma(\hat{y})) + (1 - y) \log(1 - \sigma(\hat{y})) \right]$ \tcp*{BCE Loss}
        
        $\Theta \leftarrow \Theta - \eta \nabla_{\Theta} \mathcal{L}$ \tcp*{Update parameters via AdamW}
    }
}


\end{algorithm}

\section{Experimental Setup}
\label{section_experiments}

To empirically validate our proposed framework, we design a progressive evaluation protocol that transitions from a highly controlled, single-speaker environment to large-scale benchmark datasets containing diverse, modern synthetic and adversarial attacks. After detailing the selected datasets, we cover the specific models used to extract the latent acoustic and phonetic posteriorgram features, the data preprocessing pipelines, and the network configuration and implementation details.

\subsection{Evaluation Datasets}
\label{sec:datasets}

We utilize three complementary datasets in our experiments, as summarized in Table~\ref{tab:dataset_summary}. Each dataset serves a distinct purpose.

\begin{table*}[htpb]
\centering
\caption{Utterance distribution, attack diversity, and evaluation motivation for the three datasets used in this study.}
\label{tab:dataset_summary}

\footnotesize
\renewcommand{\arraystretch}{1.15}

\begin{tabularx}{\textwidth}{
@{}
l
l
r
r
r
>{\raggedright\arraybackslash}p{4.8cm}
c
>{\raggedright\arraybackslash}p{2cm}
@{}
}
\toprule

\textbf{Dataset} &
\textbf{Partition} &
\textbf{Bonafide} &
\textbf{Spoofed} &
\textbf{Total} &
\textbf{Evaluation Motivation} &
\textbf{\#Attacks} &
\textbf{Source Dataset(s)} \\

\midrule

\multirow{3}{*}{\textbf{\shortstack[l]{LJSpeech\\(TTS-derived)}}}
& \multirow{3}{*}{\begin{tabular}[c]{@{}l@{}}
Train\\[0.45em]
Test
\end{tabular}}
& \multirow{3}{*}{\begin{tabular}[c]{@{}r@{}}
451\\[0.45em]
49
\end{tabular}}
& \multirow{3}{*}{\begin{tabular}[c]{@{}r@{}}
451\\[0.45em]
49
\end{tabular}}
& \multirow{3}{*}{\begin{tabular}[c]{@{}r@{}}
902\\[0.45em]
98
\end{tabular}}
& \multirow{3}{4.8cm}{Isolates low-level synthesis artifacts by controlling lexical and prosodic variance (same-speaker, same-text).}
& \multirow{3}{*}{1}
& \multirow{3}{2cm}{LJSpeech} \\

& & & & & & & \\
& & & & & & & \\

\midrule

\multirow{3}{*}{\textbf{\shortstack[l]{ASVspoof 2019\\(LA)}}}
& Train & 2,580 & 22,800 & 25,380
& \multirow{3}{4.8cm}{Standard benchmark evaluating generalization across multiple speakers and unseen attacks.}
& \multirow{3}{*}{19}
& \multirow{3}{2cm}{VCTK} \\

& Dev & 2,548 & 22,296 & 24,844
& & & \\

& Eval & 7,355 & 63,882 & 71,237
& & & \\

\midrule

\multirow{3}{*}{\textbf{\shortstack[l]{ASVspoof 5\\(Track 1)}}}
& Train & 18,797 & 163,560 & 182,357
& \multirow{3}{4.8cm}{In-the-wild evaluation with crowdsourced audio, neural vocoders, and adversarial filtering attacks.}
& \multirow{3}{*}{32}
& \multirow{3}{2cm}{MLS (English) / LibriVox} \\

& Dev & 31,334 & 109,616 & 140,950
& & & \\

& Eval & 138,688 & 542,086 & 680,774
& & & \\

\bottomrule
\end{tabularx}
\end{table*}

\textbf{LJSpeech (TTS-derived):} The LJSpeech corpus~\cite{ito2017lj} is a public dataset comprising approximately 13{,}100 short audio clips of a single English-speaking female reader narrating passages from seven non-fiction books, with a total duration of roughly 24 hours. The recordings were captured under quiet, near-studio conditions, with sentence-level segmentation distributed alongside the dataset; the prosody is relatively flat and the delivery follows a ``read-aloud'' style typical of audiobook narration. To establish baseline behavior of our detector under maximally controlled conditions, we use a publicly available LJSpeech-derived spoofing detection corpus~\cite{ito2017lj} that contains parallel real and synthesized utterances of this single speaker: every synthetic recording is paired with a corresponding bonafide utterance having identical content. The spoofed utterances are generated by a single attack pipeline, namely Tacotron~2~\cite{shen2018natural} as the acoustic model coupled with Parallel WaveGAN~\cite{yamamoto2020parallel} as the neural vocoder. Because both the speaker identity and the linguistic content are held fixed across the bonafide--spoof pairs, all lexical, prosodic, and speaker-related variability are effectively eliminated, and the detection models are forced to rely on synthesis artifacts introduced by this specific TTS system. We adopt the train/test partition distributed with this corpus directly; no development partition is provided.

\textbf{ASVspoof 2019:}
To evaluate the framework against a widely adopted standard, we utilize the Logical Access (LA) partition of the ASVspoof 2019 dataset~\cite{wang2020asvspoof}. The LA detection partition that we use contains 87 speakers in total (38 male, 49 female), with 20 speakers (8 male, 12 female) shared between the training and development sets and a disjoint set of 67 speakers (30 male, 37 female) in the evaluation set. The dataset encompasses a variety of Text-to-Speech (TTS) and Voice Conversion (VC) algorithms. The training and development sets feature attacks generated by 6 methods (A01 through A06). The evaluation partition contains 13 unseen attack algorithms (A07 through A19), providing a robust test of the model's ability to generalize to novel synthesis or conversion systems.

\textbf{ASVspoof 5:} To assess the performance using modern, in-the-wild threats, we evaluate on Track~1 (stand-alone spoofing and speech deepfake detection) of ASVspoof~5~\cite{wang2026asvspoof}, the latest edition of the ASVspoof challenge series. Unlike the previous ASVspoof challenge editions, which relied on studio-quality source recordings, ASVspoof~5 is built from crowdsourced, non-studio data spanning hundreds of speakers across diverse acoustic conditions. The spoofed material is generated by an extensive mix of legacy and contemporary neural TTS and VC systems together with adversarial attacks. Further, the dataset contains several codecs applied on top of both bonafide and spoofed utterances to simulate realistic transmission and storage pipelines. We utilize the full released evaluation partition (approximately 680k utterances).

\subsection{Feature Extractors}

Table~\ref{tab:model_params} lists the two frozen subcomponents indicated in Figure~\ref{fig:architecture_diagram}.

\begin{table}[htbp]
\centering
\caption{Parameter footprint of the pre-trained feature extractors.}
\label{tab:model_params}

\begin{tabular}{llc}
\toprule
Representation & Pre-trained Estimator Model & \#Params \\
\midrule
Phonetic Posteriors ($W$) 
& Wav2Vec 2.0 (Large, TIMIT fine-tuned)\footnotemark[4]
& 315{,}506{,}370 \\

Acoustic Features ($X$)
& XLS-R (300M, Multilingual)\footnotemark[5]
& 315{,}438{,}720 \\
\bottomrule
\end{tabular}

\end{table}

\footnotetext[4]{Available at: \url{https://huggingface.co/excalibur12/wav2vec2-large-lv60_phoneme-timit_english_timit-4k}} \footnotetext[5]{Available at: \url{https://huggingface.co/facebook/wav2vec2-xls-r-300m}}

\textbf{Acoustic SSL Model ($\mathcal{F}_X$):} To capture the low-level spectral and temporal artifacts generated by neural vocoders, we adopt the XLS-R (Cross-Lingual Speech Representation) model~\cite{babu2021xls} as our acoustic front-end. Specifically, we use the 300-million parameter variant of XLS-R, based on the Wav2Vec 2.0 architecture and pre-trained on 436k hours of unlabeled speech across 128 languages (including VoxPopuli, CommonVoice, BABEL, and VoxLingua107)~\cite{babu2021xls}. We extract the features from the final hidden state of the model, yielding the temporal acoustic sequence $X \in \mathbb{R}^{T_x \times 1024}$, where $T_x$ is the number of XLS-R output frames produced for a given utterance. As is standard for Wav2Vec 2.0 --derived models operating on 16~kHz audio, the convolutional feature encoder applies a total stride of 320 samples, so $F_X$ emits one 1024-dimensional embedding every 20~ms (equivalently, 50 frames per second).

\textbf{Phonetic Posteriorgram Extractor ($\mathcal{F}_W$):} To generate the phonetic posteriorgrams (PPGs), we use a pre-trained Wav2Vec 2.0 Large (lv60) model\footnotemark[6] that was made publicly available after being fine-tuned by the original authors on the English TIMIT corpus~\cite{garofolo1993timit} for frame-level phone classification. As reported by those authors, the model was optimized over the 4{,}620-utterance TIMIT training set with the Adam optimizer, achieving a phone error rate (PER) of 10.53\% on the 1{,}680-utterance core test set, a result competitive with the most accurate phone-recognition systems reported on TIMIT and approaching the natural lower bound set by inter-listener phone-classification disagreement\footnotemark[7]. We employ this model directly without further fine-tuning: each utterance is passed through the network, and the resulting logits are softmax-normalized to yield a sequence of probability distributions over the standard 61-phone TIMIT alphabet, satisfying the constraint $\sum_{i=1}^{61} w_i = 1$ at every frame. Operating on 16~kHz audio, $F_W$ emits one 61-dimensional PPG vector every 20~ms (50 frames per second), placing it on the same temporal grid as $F_X$ (Section~\ref{subsec:frontend}) and thereby making the alignment between the two streams straightforward.

\footnotetext[6]{Both $F_X$ (XLS-R) and $F_W$ (Wav2Vec~2.0 Large lv60 TIMIT-fine-tuned) share the same underlying wav2vec~2.0 transformer architecture, which is why their parameter counts in Table~\ref{tab:model_params} are nearly identical. The two extractors differ in their pre-training data ($F_X$ is pre-trained on 436{,}000 hours of multilingual speech across 128 languages, while $F_W$ is fine-tuned for English phone classification on TIMIT) and in their downstream role: $F_X$ provides contextualized acoustic representations directly from the final hidden state, while $F_W$ produces frame-level phonetic posteriors via a softmax-normalized classification head.}

\footnotetext[7]{For context, recent self-supervised systems on TIMIT typically report PER in the range of 8--10\% (e.g.,~\cite{baevski2020wav2vec}, where wav2vec 2.0 achieves a PER of 7.4\% on TIMIT, and~\cite{hsu2021hubert}, where HuBERT reports a PER of 8.5\% on the same benchmark), with human-listener phone-classification error rates also frequently cited in the 5--10\% range, placing the present model within the strong-performing tier on this benchmark.}

\subsection{Data Preprocessing}

For computational efficiency and elimination of redundant forward passes during training, the acoustic embeddings ($X$) and phonetic posteriorgrams ($W$) were pre-extracted and stored offline for all utterances across all datasets. Resampling was performed using the polyphase FIR-based implementation in \texttt{torchaudio.functional.resample}, which applies a Kaiser-windowed low-pass anti-aliasing filter prior to decimation/interpolation to suppress spectral aliasing. 

A well-known \emph{shortcut artifact} in anti-spoofing arises when synthesizers emit literal zero-valued samples in silent regions while bonafide recordings retain natural room tone, allowing detectors to exploit the resulting noise-floor discrepancy rather than learning genuine spoofing cues~\cite{chettri2020dataset,muller2021speech}. Such shortcut learning, where deep models latch onto incidental statistical regularities in the training data rather than the intended task-relevant signal, has been documented as a general phenomenon across deep learning~\cite{geirhos2020shortcut}. To mitigate this potential learning shortcut on the multi-speaker corpora, we applied an energy-based speech activity detection (SAD) algorithm to trim ambient silence from the ASVspoof~2019 LA and ASVspoof~5 datasets, ensuring that the detection models are forced to evaluate the active articulatory mechanics of the speech rather than exploiting trivial background noise discrepancies.

Once the PPGs were extracted, five structural tokens (\texttt{|}, \texttt{[UNK]}, \texttt{[PAD]}, \texttt{<s>}, \texttt{</s>}) inherent to the Wav2Vec~2.0 vocabulary were pruned from the posterior distributions. These tokens denote word boundaries (\texttt{|}), unknown/out-of-vocabulary symbols (\texttt{[UNK]}), batching-time padding (\texttt{[PAD]}), and sentence start/end markers (\texttt{<s>}, \texttt{</s>}); they are decoding-time bookkeeping symbols rather than phonetic units. We therefore exclude them so that $W$ remains a strict distribution over articulatory categories. Although in principle the activations on these tokens could carry incidental cues about synthesis pipelines, retaining them would dilute the interpretability of the downstream phone-conditional decomposition; we therefore prune them \emph{by design}. This yields the linguistically grounded phonetic dimensionality of $M = 61$, corresponding exactly to the TIMIT phone inventory listed in Table~\ref{tab:phoneme_groups}. Finally, temporal alignment between the XLS-R and PPG streams was therefore achieved on the fly by truncating both sequences to their minimum shared length, 
$T = \min(T_x, T_p)$, to absorb any minor length discrepancies arising in practice.


\subsection{Model Configurations and Baselines}
\label{configs}

\textbf{Proposed Cross-Attention Architecture:} The phoneme-guided cross-attention network accepts the temporally aligned inputs $X \in \mathbb{R}^{T \times 1024}$ and $W \in \mathbb{R}^{T \times 61}$ produced by the front-end (Section~\ref{subsec:frontend}). Inside the cross-attention block, both streams are projected to a common attention hidden dimensionality of $H = 320$: the acoustic stream is mapped from its native 1024 dimensions to $H$ through the key and value projections of Eq.~\eqref{eq:keys_values}, and the phonetic queries are constructed in $\mathbb{R}^{M \times H}$ as in Eq.~\eqref{eq:queries}. Operating on the union of features (effective dimensionality $1024 + 61 = 1085$) without projection would substantially increase the number of attention parameters and risk overfitting on the smaller training partitions (notably LJSpeech); we therefore use a moderate $H$ that compresses the acoustic stream while preserving enough capacity for the 61 phonetic anchors. We fix $H = 320$ as a default that balances representational capacity against parameter count and visual RAM in our graphics processing unit (GPU) computational environment cost at our adopted batch size; a systematic sweep over $H$ is left to future work.

Our cross-attention module is configured with a single attention head ($h = 1$). Many transformer-based systems use $h \gg 1$ on the premise that distinct heads can specialize on complementary aspects of the input. The use of multiple heads often yields improved predictive performance improved performance~\cite{vaswani2017attention}. We purposefully restrict the architecture to a single head, however, because of the central role that interpretability plays in this work: under $h = 1$ each phonetic anchor in $Q$ attends to the acoustic stream through one shared attention map, so the per-row content of $A_{\text{out}}[m]$ admits a direct, unambiguous reading as the acoustic evidence associated with phone class $m$ (as developed in Section~\ref{subsec:crossattn}). Splitting attention across multiple heads would distribute the alignment between phonetic anchors and acoustic frames across several non-comparable subspaces, complicating the phone-conditional interpretation that motivates the entire framework. This trade-off, accepting a small potential loss in raw discriminative capacity in exchange for a clean phone-aligned representation, is intentional, and our experiments on ASVspoof~5 (Section~\ref{results}) confirm that the single-head configuration reaches competitive EER.

Following the cross-attention layer, the aggregated phoneme representations are passed through a binary classifier consisting of a linear projection ($320 \rightarrow 256$), a ReLU activation, dropout with retention probability $p = 0.2$ (i.e., each unit is independently zeroed with probability $0.2$ during training), and a final linear layer ($256 \rightarrow 1$). To evaluate the aggregation of phoneme-specific evidence, we consider two pooling variants of the cross-attention model. Under the \emph{mean pooling} variant (Eq.~\eqref{eq:mean_pool}), the phonetic representations produced by the cross-attention module are averaged uniformly across the $M$ rows, corresponding to the choice $w_i = 1/M$ in Eq.~\eqref{eq:final_factorization}. Under the \emph{learned weighted pooling} variant (Eqs.~\eqref{eq:pool_logits}--\eqref{eq:weighted_pool}), a learned linear scoring layer $W_p \in \mathbb{R}^{H \times 1}$ maps each row $A_{\text{out}}[m]$ to a scalar logit, which is then softmax-normalized over the $M$ phonetic classes to yield the data-driven weights $\alpha_m$. The utterance embedding is the $\alpha$-weighted sum of the phonetic rows, $E_{\text{utt}} = \sum_{m=1}^{M} \alpha_m A_{\text{out}}[m].$
Because $\alpha$ is produced through an explicit, supervised, end-to-end-trained projection, the model learns to assign mass to those phonetic classes whose row $A_{\text{out}}[m]$ most consistently separates bonafide from spoofed speech under the binary cross-entropy objective. This makes the per-class $\alpha_m$ directly inspectable at inference time and provides the structural interface used for the targeted phonetic masking experiments described in Section~\ref{ablation_protocol}.

\textbf{Self-Attention Baselines:} To quantify the added value of our dual-stream cross-attention framework, we implemented standard self-attention baselines, which have been widely adopted in recent anti-spoofing systems such as RawNet2~\cite{tak2021end} and AASIST~\cite{jung2022aasist}. These baselines mirror the proposed model in hidden dimensionality ($H = 320$), number of attention heads ($h = 1$), and downstream classifier architecture, but differ in that they operate on an isolated single feature stream. The queries, keys, and values are derived from one and the same input sequence: in the \textit{PPG-only} baseline, $Q$, $K$, and $V$ are all linear projections of the phonetic posteriorgram $W$, testing the semantic separability achievable from phonetic information alone; in the \textit{XLSR-only} baseline, $Q$, $K$, and $V$ are all linear projections of the acoustic embedding $X$, testing the acoustic separability achievable without any phonetic anchoring. No mixing between the two streams occurs at any stage of these baselines, which is the architectural property that the proposed cross-attention model is designed to add.

\subsection{Training Protocol and Evaluation Metrics}
\label{eval_metrics}

All models were implemented in PyTorch. The networks were optimized end-to-end using binary cross-entropy loss, computed in its numerically stable logit form (\texttt{BCEWithLogitsLoss}, which fuses a sigmoid activation with the standard binary cross-entropy computation in a single operation to avoid floating-point underflow). We used the AdamW optimizer~\cite{loshchilov2017decoupled} with a learning rate of $10^{-4}$ and a weight decay coefficient of $10^{-4}$.

To accommodate video random access memory constraints while maximizing throughput, the cross-attention models were trained with a batch size of 4 and the simpler self-attention baselines with a batch size of 16. All training runs were initialized with a fixed random seed for reproducibility. Training duration was adapted to dataset scale and computational budget: models evaluated on the smaller LJSpeech corpus were trained for up to 100 epochs, while those trained on the substantially larger ASVspoof~2019 and ASVspoof~5 datasets were capped at 20 epochs. By inspecting the training and validation loss trajectories on ASVspoof~2019, we verified that this 20-epoch cap was sufficient by inspecting the training and validation loss trajectories on ASVspoof~2019 (\ref{app:convergence}); both curves had largely flattened well before the cap, with validation EER reaching its minimum within the trained window. Across all configurations, the final model checkpoint was selected based on optimal validation-set performance.

We evaluate all models using two complementary metrics. The primary metric is the equal error rate (EER), defined as the operating point on the ROC curve at which the miss rate $P_{\text{miss}}$ (bonafide utterances incorrectly classified as spoof) equals the false-alarm rate $P_{\text{fa}}$ (spoof utterances incorrectly classified as bonafide). To complement EER, which fixes equal cost on the two error types, we also report the minimum normalized detection cost function (\textit{minDCF})~\cite{delgado2024asvspoof}, given by

\begin{equation}
\text{minDCF} = \min_{\tau}
\frac{
C_{\text{miss}} P_{\text{miss}}(\tau) P_{\text{tgt}}
+
C_{\text{fa}} P_{\text{fa}}(\tau) (1 - P_{\text{tgt}})
}{
\min\!\left(
C_{\text{miss}} P_{\text{tgt}},
\;
C_{\text{fa}} (1 - P_{\text{tgt}})
\right)
},
\end{equation}
with $C_{\text{miss}} = C_{\text{fa}} = 1$ and target prior $P_{\text{tgt}} = 0.05$. The denominator is the cost of an uninformative reference classifier that ignores the input, by either always accepting or rejecting the input (whichever gives a lower cost). $\text{minDCF} = 1$ therefore corresponds to a system that performs no better than this naive baseline, and $\text{minDCF} = 0$ to a perfect detector. 



To quantify the sampling uncertainty of both reported metrics, we apply non-parametric bootstrap resampling~\cite{efron1992bootstrap}, drawing 1{,}000 bootstrap resamples (with replacement) of the per-utterance scores and reporting 95\% confidence intervals as the 2.5--97.5 percentile range of the resulting bootstrap distribution. This approach is preferred over parametric alternatives because it does not require any assumption about the underlying score distribution. This is particularly well suited to EER and minDCF whose sampling distributions are typically non-Gaussian.

\subsection{Targeted Phonetic Ablation Protocol}
\label{ablation_protocol}

To empirically isolate which articulatory groups carry the most discriminative information, we conduct a targeted ablation study on the ASVspoof~2019 LA partition using the proposed cross-attention model with weighted pooling. The seven articulatory groups defined in Table~\ref{tab:phoneme_groups} (\emph{Vowels}, \emph{Stops}, \emph{Affricates}, \emph{Fricatives}, \emph{Nasals}, \emph{Semivowels}, \emph{Other}) serve as the units of ablation. For each group $g$, we train and evaluate an \emph{single-group restriction} configuration in which the model is restricted to the phones in group $g$: all rows of $A_{\text{out}}$ outside group $g$ (defined per Eq.~\eqref{eq:Aout} in Section~\ref{subsec:crossattn}) are suppressed before the weighted sum, and only the surviving rows contribute to the utterance embedding $E_{\text{utt}}$ (defined per Eq.~\eqref{eq:weighted_pool} in Section~\ref{subsec:backend}). By limiting the model's phonetic budget to a single articulatory group at a time, this design directly probes the discriminative information each group carries on its own.

We note an important caveat to the interpretation of these ablation results. Because the seven articulatory groups occur at substantially different frequencies in natural speech, the \emph{single-group restriction} configurations are not directly comparable in terms of the effective training signal available to each group: frequently occurring groups such as vowels naturally contribute more usable frames per training utterance than rarer groups such as affricates. The resulting differences in Eval EER across groups therefore entangle two distinct factors: the discriminative power inherent to each articulatory category and the amount of training signal available to fit the corresponding restricted model. A fully controlled comparison would require equalizing the per-group training-frame counts across configurations. We therefore frame the resulting EER differences as indicative of the relative discriminative content of each group rather than as a definitive ranking.

We realize this restriction in two complementary ways. In \emph{score masking}, the pooling logits $L_p[m]$ of all phones outside group $g$ are set to $-\infty$ before the softmax in Eq.~\eqref{eq:alpha}. This way, the corresponding $\alpha_m$ are exactly zero and the surviving group's weights renormalize over its phones. In \emph{vector zeroing}, the softmax weights $\alpha$ are computed normally over all $M$ phones, but the rows $A_{\text{out}}[m]$ of phones outside group $g$ are zeroed prior to the weighted sum. The two enforcement schemes restrict the model to the same phonetic budget but differ in how the original attention mass is treated: score masking forces the surviving group to absorb the full probability budget, whereas vector zeroing preserves the original weight distribution and simply withholds the masked phones' contribution.

To match training and evaluation conditions, a separate copy of the cross-attention model is trained from scratch for every group under each enforcement scheme using a correspondingly filtered PPG stream, with all hyperparameters held fixed at the values described in Section~\ref{configs}, \emph{with one deliberate exception}: the ablation models are trained with $h=8$ attention heads, rather than the $h=1$ used by the full model. The reason for this change is methodological rather than architectural. In the full model, $h=1$ is adopted for \textbf{interpretability}, so that each phonetic anchor in $Q$ attends to the acoustic stream through one shared attention map whose contents are directly inspectable. In the ablation models, however, the goal is not to inspect attention maps but to assess whether each phonetic group, on its own, carries enough \textbf{discriminative information} to support detection. Restricting the model to a single group already imposes a substantial capacity bottleneck on the architecture; pairing that restriction with a single-head attention block would make it difficult to disentangle a poor result caused by \emph{this group genuinely lacks discriminative information} from a poor result caused by \emph{the single-head attention block could not extract the available information under such a narrow phonetic budget}. Allowing $h=8$ heads gives each ablation model a fair chance to extract whatever detectable spoofing traits remain in its restricted budget, and any residual EER gap across groups can then be attributed to the phonetic content itself rather than to an attention-capacity bottleneck. EER and minDCF, together with their bootstrap 95\% confidence intervals, are computed on the ASVspoof~2019 LA evaluation partition.

\section{Experimental Results}
\label{results}

We organize the results by dataset.

\subsection{LJSpeech} \label{subsec:results_ljspeech}

Table~\ref{tab:LJ_model_results} summarizes detection performance on the controlled, same-speaker, same-text LJSpeech (TTS-derived) corpus. A clear progression emerges across the three systems. The PPG-only self-attention baseline performs at near-chance level, with test EER values of $42.85\%$ (mean pooling) and $40.81\%$ (weighted pooling) and corresponding minDCF values close to the upper bound of $1.0$. This outcome is consistent with the role of phonetic posteriorgrams as content-oriented representations: PPGs are designed to encode \emph{what was said} while treating speaker-, channel-, and realization-level variation as \emph{nuisance factors} to be suppressed. From a detection perspective, however, these very nuisance factors are arguably expected to contain the bulk of the spoofing-relevant signal, since they encode the fine-grained, non-linguistic detail in which generators are most likely to introduce artifacts. The LJSpeech construction further amplifies this effect, as the same-speaker, same-text paradigm removes any residual lexical or prosodic discrepancy that might otherwise leak into the phonetic stream and provide a discriminative shortcut. 

Replacing the phonetic stream with the XLS-R acoustic stream improves performance substantially, with test EER dropping to $22.44\%$ for mean pooling and $17.34\%$ for weighted pooling. This result aligns with prior work that has used XLS-R as a strong acoustic backbone for anti-spoofing ~\cite{babu2021xls}, confirming that the self-supervised acoustic embeddings retain a much richer mixture of cues, including fine-grained spectral and prosodic-realization detail, that is directly relevant to distinguishing bonafide from synthesized speech. We note that the absolute EER values on LJSpeech are markedly higher than those typically reported on standard anti-spoofing benchmarks. This gap may reflect broader domain-generalization considerations, as LJSpeech differs substantially from the multi-speaker, multi-attack training distributions on which most countermeasures are benchmarked; a more definitive characterization of which factors drive this gap would require a dedicated study.

The proposed PPG~+~XLS-R cross-attention model achieves the strongest performance, with a test EER of $12.24\%$ under both pooling variants and a corresponding minDCF of $0.77$. The fact that PPGs are weak on their own yet produce a clear additional improvement when paired with XLS-R through cross-attention indicates that phonetic structure functions here as a \emph{phonetic alignment frame}: it organizes the acoustic evidence into phone-aligned slots that the cross-attention block can then probe selectively, following the decomposition of Eq.~\eqref{eq:final_factorization}.

Regardless of the detection model, the two pooling variants are broadly comparable. For the cross-attention model, the two pooling variants yield identical test EER and minDCF, while for the XLSR-only baseline weighted pooling yields a modest EER improvement of roughly five percentage points. We adopt weighted pooling as the reference configuration for the subsequent analyses, not because it is decisively stronger on this dataset, but because it produces an explicit, inspectable set of per-phone weights $\alpha_m$ (Eq.~\eqref{eq:alpha}). These weights serve as the structural interface that enables both the per-phone importance rankings reported next and the targeted-masking ablation introduced later in Section~\ref{subsec:results_ablation}.

\begin{table}[h!]
\centering
\caption{Detection performance on LJSpeech (TTS-derived) corpus across the PPG-only and XLS-R-only self-attention baselines and the proposed PPG~+~XLS-R cross-attention model, under mean and weighted pooling. Confidence intervals are shown as \emph{[lower, upper]} values. The intervals are noticeably wide on LJSpeech, particularly for the weaker baselines, because the test partition contains only 98 utterances, which limits the stability of the bootstrap sampling distribution. Some upper bounds for the PPG-only baseline EER exceed the theoretical maximum of $50\%$ because the percentile bootstrap is unconstrained by the metric's natural bound.}
\label{tab:LJ_model_results}

\resizebox{\textwidth}{!}{%
\begin{tabular}{l l l c c c c}
\hline
Features & Model & Pooling & Train Accuracy [CI]\% $\uparrow$ & Test Accuracy [CI]\% $\uparrow$ & Test EER [CI]\% $\downarrow$ & Test minDCF [CI]$\downarrow$ \\
\hline

\multirow{2}{*}{PPG}
& \multirow{2}{*}{Self-Attention}
& Mean
& 50.11 [46.56, 53.22] & 50.00 [39.80, 60.20] & 42.85 [34.04, 54.16] & 1.00 [0.89, 1.00] \\
& & Weighted
& 50.44 [47.23, 53.77] & 50.00 [39.80, 60.20] & 40.81 [32.65, 53.48] & 0.97 [0.90, 1.00] \\

\multirow{2}{*}{XLSR}
& \multirow{2}{*}{Self-Attention}
& Mean
& 70.18 [67.18, 73.28] & 77.55 [64.29, 84.69] & 22.44 [15.55, 35.00] & 0.91 [0.67, 0.98] \\
& & Weighted
& 72.51 [69.18, 75.83] & 82.65 [69.39, 88.78] & 17.34 [11.11, 29.63] & 0.95 [0.75, 1.00] \\

\multirow{2}{*}{PPG + XLSR}
& \multirow{2}{*}{Cross-Attention}
& Mean
& 90.69 [88.69, 92.68] & 87.76 [79.59, 93.88] & 12.24 [6.38, 20.40] & 0.77 [0.28, 0.87]  \\
& & Weighted
& 87.14 [84.81, 89.47] & 87.76 [77.55, 92.86] & 12.24 [7.40, 22.22] & 0.77 [0.21, 0.88]  \\

\hline
\end{tabular}%
}
\end{table}

\begin{figure}[h!]
    \centering
\includegraphics[width=\linewidth]{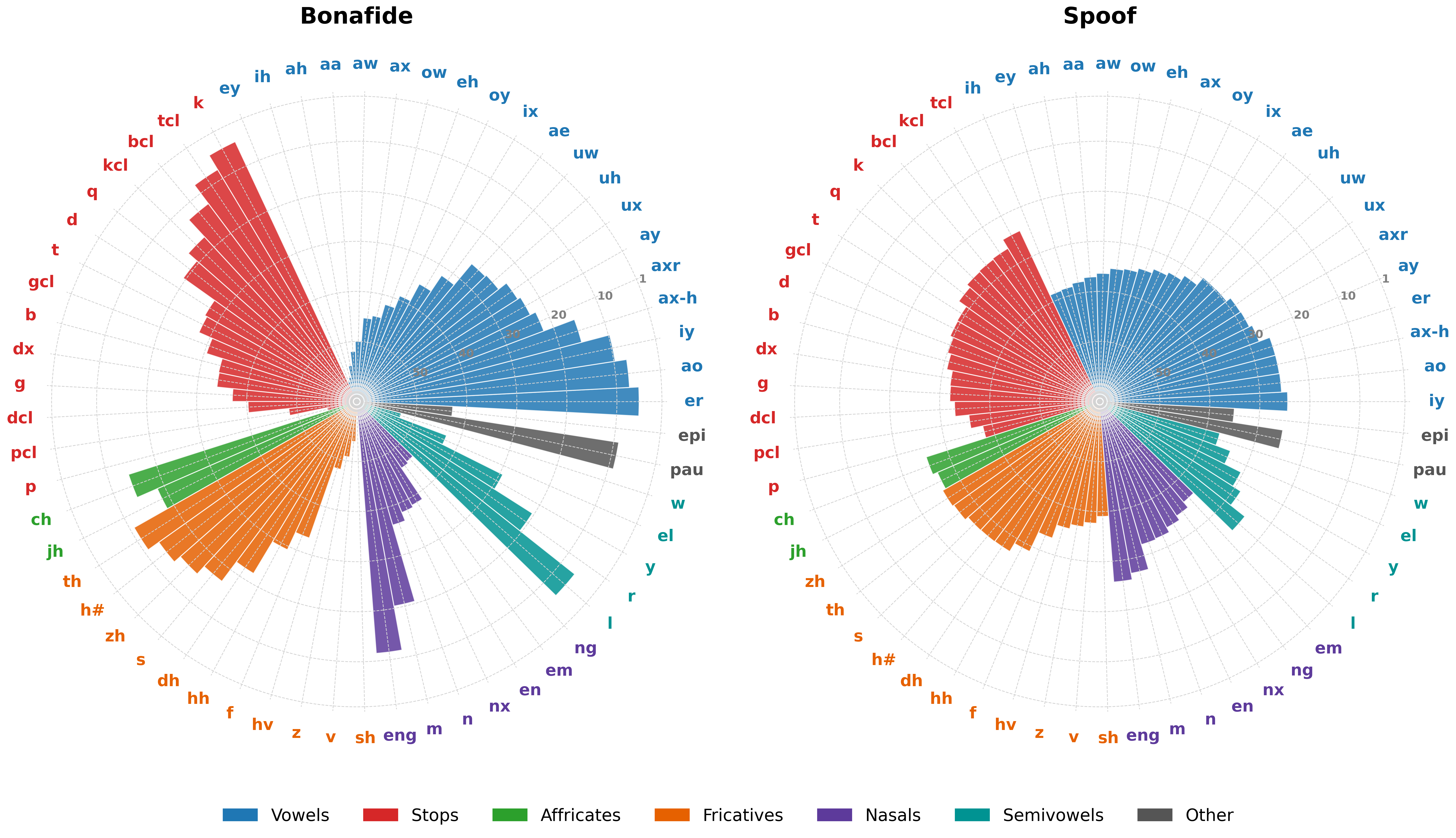}
\caption{Per-phone importance rankings of the proposed cross-attention model with weighted pooling on LJSpeech, derived by ranking the learned weights $\alpha_m$ per utterance and averaging across the bonafide (left) and spoof (right) classes. Bar length encodes average importance: the longest bars correspond to the most strongly attended phones, with stops, semivowels, and fricatives consistently dominating the top-ranked positions across both classes.}
    \label{fig:LJSpeech_figure}
\end{figure}

The phone-importance pattern produced by the weighted cross-attention model is shown in the circular bar plots in Figure~\ref{fig:LJSpeech_figure}. For each test utterance, the per-phone $\alpha_m$ values were first ranked. The ranks were then averaged separately across the bonafide and spoof classes to produce the two circular panels; phones are color-coded by the seven articulatory groups following Table~\ref{tab:phoneme_groups}. Three immediate observations stand out.

\begin{enumerate}
\item The bonafide and spoof rankings are broadly similar at the broad phone groups, which is an expected consequence of LJSpeech's same-speaker, same-text design: both classes contain effectively the same phonemic content. Hence, any divergence in the rankings must reflect class-conditional differences in how strongly the model relies on each phonetic anchor rather than differences in phonetic composition;

\item The model concentrates importance non-uniformly across articulatory groups. The most strongly attended phones are dominated by stops (\texttt{k}, \texttt{tcl}, \texttt{pcl}, \texttt{bcl}, \texttt{kcl}), a handful of high-rank vowels and semivowels (\texttt{er}, \texttt{iy}, \texttt{l}), and several fricatives (\texttt{s}, \texttt{th}, \texttt{zh}), with the silence-related symbols (\texttt{pau}) also appearing near the top. By contrast, peripheral phones such as several mid- and back-vowels (\texttt{ih}, \texttt{ey}, \texttt{ah}), most nasals (\texttt{m}, \texttt{n}, \texttt{em}, \texttt{en}, \texttt{ng}), and two affricates (\texttt{ch}, \texttt{jh}) are consistently down-weighted;

\item The magnitudes of the rank values differ markedly between the two classes: bonafide ranks span a wide range ($\sim 5$ to $\sim 59$), whereas spoof ranks compress into a narrower band ($\sim 24$ to $\sim 39$). This compression on the spoof side indicates that the model spreads its attention more uniformly across phonetic categories when scoring synthetic speech, while on bonafide speech it commits more decisively to a small set of strongly attended phones.
\end{enumerate}

Taken together, these observations identify \emph{stops}, \emph{fricatives}, and \emph{semivowels} as the articulatory groups that carry the dominant discriminative signal under this model, directly supporting the central premise of Section~\ref{related_works}: detection is most effective when driven by the phonetic categories where generative models struggle to faithfully reproduce articulatory dynamics.

\subsection{ASVspoof 2019} \label{subsec:results_asvspoof2019}

Table~\ref{tab:asvs19_table} reports detection performance on the ASVspoof~2019 LA partition, where the model is exposed to multi-speaker variability, including $13$ unseen TTS and VC attack algorithms (A07--A19) in the evaluation set. Two observations stand out by comparison with LJSpeech. First, the single-stream XLS-R self-attention baseline is now a remarkably strong system, attaining $7.43\%$ Eval EER with mean pooling and $6.95\%$ with weighted pooling, which is an improvement of more than $10$ percentage points relative to the same baseline on LJSpeech. These results are consistent with prior work that has used wav2vec 2.0 and XLS-R as anti-spoofing backbones on this dataset~\cite{tak2022automatic}, where similar EER ranges have been reported under comparable single-stream setups. The most plausible explanation for the discrepancy between the two datasets is that the multi-speaker, multi-attack training distribution of ASVspoof~2019 simply provides far more discriminative signal to a model that operates directly on rich self-supervised acoustic embeddings, and that the dataset's training set is large enough to fully exploit XLS-R's expressive capacity. Second, the proposed PPG~+~XLS-R cross-attention model achieves $9.80\%$ and $7.46\%$ Eval EER under mean and weighted pooling, respectively, with the weighted variant closing essentially all of the gap to the XLS-R-only baseline. The PPG~+~XLS-R model also produces comparable minDCF values to the baseline ($0.26$ vs.\ $0.24$). 

\begin{table}[h!]
\centering
\caption{Detection performance on the ASVspoof 2019 LA partition across the XLS-R-only self-attention baseline and the proposed PPG~+~XLS-R cross-attention model, under mean and weighted pooling. Confidence intervals are shown as \emph{[lower, upper]} values.}
\label{tab:asvs19_table}

\resizebox{\textwidth}{!}{%
\begin{tabular}{l l l c c c c}
\hline
Features & Model & Pooling & Dev EER [CI]\% $\downarrow$ & Dev minDCF [CI] $\downarrow$ & Eval EER [CI]\% $\downarrow$ & Eval minDCF [CI] $\downarrow$ \\
\hline

\multirow{2}{*}{XLSR}
& \multirow{2}{*}{Self-Attention}
& Mean
& 0.32 [0.24, 0.44] & 0.02 [0.01, 0.03] & 7.43 [7.19, 7.63] & 0.25 [0.22, 0.27] \\
& & Weighted
& 0.39 [0.30, 0.49] & 0.02 [0.01, 0.04] & 6.95 [6.75, 7.17] & 0.24 [0.21, 0.26] \\

\hline

\multirow{2}{*}{PPG + XLSR}
& \multirow{2}{*}{Cross-Attention}
& Mean
& 0.48 [0.34, 0.66] & 0.03 [0.01, 0.04] & 9.80 [9.58, 10.04] & 0.27 [0.24, 0.29] \\
& & Weighted
& 0.47 [0.35, 0.69] & 0.04 [0.02, 0.07] & 7.46 [7.25, 7.69] & 0.26 [0.23, 0.28] \\

\hline
\end{tabular}%
}
\end{table}







The contrast between mean and weighted pooling is more pronounced here than on LJSpeech. For the cross-attention model, the weighted variant improves test EER from $9.80\%$ to $7.46\%$, whereas on LJSpeech the two variants produced essentially identical EERs. The improvement is qualitatively consistent with the theoretical prediction of Section~\ref{section_4}: when the training distribution contains genuine cross-phone variation in discriminative utility, a uniform pooling prior actively dilutes the per-phone evidence by treating all phonetic anchors as equally important, while the learned $\alpha_m$ weights allow the network to concentrate evidence on the phones where bonafide and spoof distributions diverge most strongly. We retain weighted pooling as the reference configuration for the subsequent analyses and ablations.

\begin{figure}[h!]
    \centering
\includegraphics[width=\linewidth]{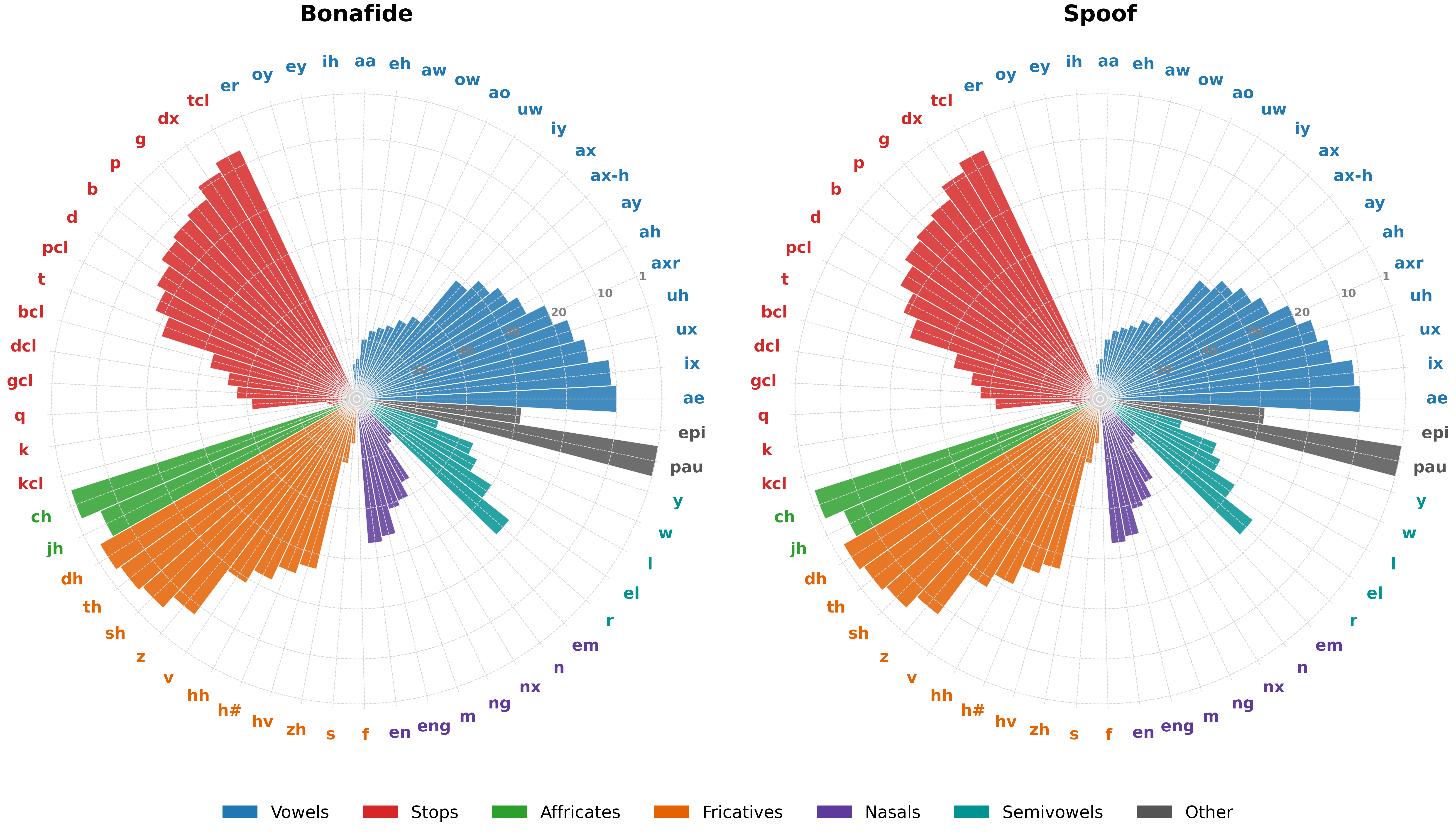}
 \caption{Per-phone importance rankings of the proposed cross-attention model with weighted pooling on ASVspoof 2019, derived by ranking the learned weights $\alpha_m$ per utterance and averaging across the bonafide (left) and spoof (right) classes. Bar length encodes average importance: the longest bars correspond to the most strongly attended phones, with stops, fricatives, and affricates consistently dominating the top-ranked positions across both classes.}
    \label{fig:ASVS19}
\end{figure}

The phone-importance pattern produced by the weighted cross-attention model on the ASVspoof~2019 LA evaluation set is shown in Figure~\ref{fig:ASVS19}. Three observations distinguish ASVspoof 2019 from LJSpeech.

\begin{enumerate}
\item The bonafide and spoof rankings are now almost identical at the phone level. This near-perfect coincidence indicates that on ASVspoof~2019, the cross-attention model has converged on a single, class-agnostic phone-importance ordering, learning \emph{where to look} on the acoustic signal but distinguishing bonafide from spoof in the per-phone content of $A_{\text{out}}[m]$ rather than in the pooling weights $\alpha_m$;

\item The top-ranked phones are dominated by the \texttt{Other} category (\texttt{pau}, \texttt{epi}), affricates (\texttt{ch}, \texttt{jh}), fricatives (\texttt{dh}, \texttt{th}, \texttt{sh}, \texttt{z}), and several stops (\texttt{tcl}, \texttt{p}, \texttt{b}, \texttt{ux}, \texttt{d}, \texttt{pcl}). This is broadly consistent with the LJSpeech pattern in which stops and fricatives dominated the rankings, but with a stronger presence of affricates and the silence/closure-related symbols. The shift toward affricates and silence-related phones may reflect the broader sensitivity of the cross-attention mechanism to discontinuities at segment boundaries---precisely the points where affricate closure--burst--turbulence sequences and silence--onset transitions occur;

\item Vowels are conspicuously down-ranked. Most front, mid, and back vowels (\texttt{eh}, \texttt{ih}, \texttt{ey}, \texttt{ao}, \texttt{aw}, \texttt{oy}, \texttt{er}) appear below rank~40, and the most strongly down-weighted phones overall are vowels and rhotic/semivowel symbols.
\end{enumerate}

Read together with the LJSpeech findings, this convergence on \emph{stops, fricatives, affricates, and silence-related closures} as the consistently top-attended groups and vowels as the consistently down-weighted group provides direct empirical support for the theoretical claim of Section~\ref{section_4}: discriminative power concentrates on the articulatory categories that generative models struggle to reproduce faithfully (transients, turbulence, anti-resonance, and silence-boundary dynamics) rather than on those they synthesize well (periodic, formant-driven vowels). The targeted-masking ablation in Section~\ref{subsec:results_ablation} further reinforces this pattern.

\subsection{Phoneme-Group Ablation on ASVspoof 2019 LA} \label{subsec:results_ablation}

To empirically test which articulatory groups carry the dominant discriminative signal under the proposed framework, we run the targeted \emph{single-group restriction} ablation protocol of Section~\ref{ablation_protocol} on ASVspoof~2019~LA. For each of the seven articulatory groups, a separate cross-attention model with weighted pooling is trained from scratch and forced to rely exclusively on the corresponding phones. Results under both enforcement schemes (\emph{score masking} and \emph{vector zeroing}) are reported in Table~\ref{tab:phoneme_ablation_sorted}, sorted by EER on the evaluation set. The reader is reminded that these ablation models use $h=8$ attention heads, in contrast to the $h=1$ configuration of the full unrestricted model (Section~\ref{configs}). Our analysis focuses on the \emph{relative ordering} of groups within Table~\ref{tab:phoneme_ablation_sorted}, which is the quantity the ablation is designed to probe.

The relative ranking of groups according to the EER calculated on the evaluation set, is essentially identical under the two ablation schemes and largely consistent with the per-phone importance pattern depicted in Figure~\ref{fig:ASVS19}. \emph{Stops} are the strongest single group: a model restricted to stops alone reaches roughly $6.6\%$ EER under both enforcement schemes. Remarkably, this is comparable to the performance of the full unrestricted model and substantially better than the other group-restricted variants. This indicates that the discriminative information concentrated in stops is sufficient on its own to support detection at a similar operating range as the full phonetic budget. \emph{Affricates}, \emph{nasals}, \emph{fricatives}, and the \emph{other} category (silence-related closures, \texttt{pau} and \texttt{epi}) form a tightly clustered second tier with Eval EERs in the $6.8\%$--$7.1\%$ range, while \emph{semivowels} ($\sim 7.51\%$) and especially \emph{vowels} ($\sim 8.18\%$--$8.21\%$) trail well behind. The relative ordering of discriminative ability — stops $>$ affricates $\approx$ nasals $\approx$ fricatives $\approx$ other $>$ semivowels $>$ vowels — mirrors the directionality of the phone-importance plot (Figure~\ref{fig:ASVS19}), where the top-ranked phones are dominated by silence/closure (\texttt{pau}), affricates (\texttt{ch}, \texttt{jh}), fricatives (\texttt{dh}, \texttt{th}, \texttt{sh}, \texttt{z}), and stops (\texttt{tcl}, \texttt{p}, \texttt{b}, \texttt{d}), while vowels and rhotic semivowels (\texttt{er}, \texttt{oy}, \texttt{ey}, \texttt{ih}) cluster at the bottom of the ranking. Score masking and vector zeroing produce nearly identical numerical outcomes for each group, with differences below $0.2\%$ Eval EER in every row; this insensitivity to the enforcement scheme indicates that the result reflects the discriminative content of the phonetic group itself rather than an artifact of how the remaining attention mass is redistributed across the surviving phones.

These findings empirically confirm the central prediction of Section~\ref{related_works}: detection accuracy is concentrated in the articulatory groups where generative models struggle to faithfully reproduce production physics, for example, transients and closures (stops, affricates), turbulent broadband noise (fricatives), anti-resonance (nasals), and silence-floor reconstruction (other); while groups that vocoders synthesize well, namely the periodic, formant-driven vowels and the gliding semivowels, contribute the least discriminative information. The systematic ordering across the seven groups, together with its agreement with the independently obtained $\alpha_m$ rankings of Figure~\ref{fig:ASVS19}, provides two convergent lines of evidence that phonetic-group identity is a first-order determinant of detection performance. This validates both the theoretical decomposition derived in Eq.~\eqref{eq:bayes_proportional} (per-phone divergence, not occurrence frequency, governs discriminability) and the architectural choice of an explicit phone-aligned bottleneck (Section~\ref{subsec:backend}) that enables this kind of targeted analysis.

\begin{table*}[h!]
\centering
\caption{Phoneme group ablation results on the ASVspoof~2019 LA evaluation partition, sorted by Eval EER within each enforcement scheme (Score Masking and Vector Zeroing). For each of the seven articulatory groups, a separate cross-attention model with weighted pooling is trained from scratch and restricted to the phones in that group. The ablation models use $h=8$ attention heads (in contrast to the $h=1$ configuration of the full unrestricted model in Table~\ref{tab:asvs19_table}). Confidence intervals are shown as \emph{[lower, upper]} values from the non-parametric bootstrap. Cell background shading provides a quick visual cue: greener shades indicate lower EER and minDCF values (better detection); redder shades indicate higher values (worse detection). The shading is normalized within each metric column to highlight the relative ordering across groups.}
\label{tab:phoneme_ablation_sorted}

\small
\setlength{\tabcolsep}{3pt}

\resizebox{\textwidth}{!}{
\begin{tabular}{llcccc}
\hline
Method & Selected Group & Dev EER [CI]\% $\downarrow$ & Dev minDCF [CI] $\downarrow$ & Eval EER [CI]\% $\downarrow$ & Eval minDCF [CI] $\downarrow$ \\
\hline

\multirow{7}{*}{Score Masking}
& stops      
& \cellcolor{green!25}0.3925 [0.2918, 0.5200] 
& \cellcolor{red!25}0.0314 [0.0085, 0.0465] 
& \cellcolor{green!40}6.6185 [6.3710, 6.8421] 
& \cellcolor{yellow!25}0.2237 [0.2017, 0.2404] \\

& affricates 
& \cellcolor{yellow!25}0.3319 [0.2401, 0.5092] 
& \cellcolor{orange!20}0.0285 [0.0077, 0.0370] 
& \cellcolor{green!30}7.0176 [6.7915, 7.2345] 
& \cellcolor{yellow!30}0.2272 [0.2002, 0.2446] \\

& nasals     
& \cellcolor{yellow!30}0.3543 [0.2692, 0.4650] 
& \cellcolor{yellow!25}0.0237 [0.0064, 0.0439] 
& \cellcolor{yellow!20}7.0292 [6.8245, 7.2716] 
& \cellcolor{green!25}0.2220 [0.2042, 0.2372] \\

& fricatives 
& \cellcolor{orange!20}0.3812 [0.3011, 0.5094] 
& \cellcolor{green!35}0.0189 [0.0073, 0.0350] 
& \cellcolor{yellow!30}7.0928 [6.8219, 7.3041] 
& \cellcolor{orange!25}0.2318 [0.2077, 0.2511] \\

& other      
& \cellcolor{yellow!30}0.3532 [0.2688, 0.4937] 
& \cellcolor{yellow!30}0.0260 [0.0116, 0.0320] 
& \cellcolor{yellow!30}7.0972 [6.8668, 7.3135] 
& \cellcolor{green!30}0.2194 [0.2007, 0.2333] \\

& semivowels 
& \cellcolor{yellow!25}0.3454 [0.2514, 0.4413] 
& \cellcolor{green!30}0.0203 [0.0071, 0.0219] 
& \cellcolor{orange!25}7.5076 [7.2825, 7.7358] 
& \cellcolor{yellow!30}0.2266 [0.2004, 0.2442] \\

& vowels     
& \cellcolor{yellow!25}0.3454 [0.2676, 0.4978] 
& \cellcolor{green!30}0.0214 [0.0080, 0.0378] 
& \cellcolor{red!30}8.2121 [7.9391, 8.4697] 
& \cellcolor{yellow!25}0.2262 [0.2015, 0.2420] \\

\hline

\multirow{7}{*}{Vector Zeroing}
& stops      
& \cellcolor{green!25}0.3925 [0.2918, 0.5217] 
& \cellcolor{red!25}0.0312 [0.0086, 0.0462] 
& \cellcolor{green!40}6.6372 [6.3879, 6.8595] 
& \cellcolor{yellow!25}0.2249 [0.2025, 0.2411] \\

& affricates 
& \cellcolor{yellow!30}0.3532 [0.2455, 0.5121] 
& \cellcolor{orange!20}0.0296 [0.0091, 0.0454] 
& \cellcolor{green!30}6.8117 [6.6253, 7.0660] 
& \cellcolor{orange!25}0.2307 [0.2045, 0.2518] \\

& other      
& \cellcolor{orange!20}0.3723 [0.2872, 0.4794] 
& \cellcolor{orange!20}0.0297 [0.0116, 0.0455] 
& \cellcolor{green!25}6.8940 [6.6777, 7.0834] 
& \cellcolor{orange!30}0.2364 [0.2144, 0.2488] \\

& nasals     
& \cellcolor{yellow!25}0.3409 [0.2678, 0.4572] 
& \cellcolor{yellow!25}0.0238 [0.0064, 0.0443] 
& \cellcolor{yellow!30}7.0428 [6.8160, 7.2747] 
& \cellcolor{green!30}0.2243 [0.2047, 0.2391] \\

& fricatives 
& \cellcolor{orange!20}0.3812 [0.3043, 0.5019] 
& \cellcolor{green!35}0.0188 [0.0075, 0.0350] 
& \cellcolor{yellow!30}7.0928 [6.8173, 7.3066] 
& \cellcolor{orange!25}0.2332 [0.2089, 0.2519] \\

& semivowels 
& \cellcolor{green!25}0.3274 [0.2516, 0.4443] 
& \cellcolor{green!30}0.0215 [0.0070, 0.0230] 
& \cellcolor{orange!25}7.5187 [7.2834, 7.7356] 
& \cellcolor{yellow!30}0.2272 [0.2027, 0.2466] \\

& vowels     
& \cellcolor{yellow!25}0.3454 [0.2648, 0.5000] 
& \cellcolor{green!30}0.0215 [0.0081, 0.0379] 
& \cellcolor{red!30}8.1838 [7.9185, 8.4568] 
& \cellcolor{yellow!25}0.2267 [0.2019, 0.2424] \\

\hline
\end{tabular}
}
\end{table*}

\subsection{ASVspoof 5} \label{subsec:results_asvspoof5}

Table~\ref{tab:asvs5_results} reports detection performance on Track~1 of ASVspoof~5, where the detection model is subjected to crowdsourced recordings, hundreds of speakers, contemporary neural TTS and VC attacks, adversarial perturbations, and multiple codec degradations applied to both bonafide and spoofed material. The XLS-R self-attention baseline attains $8.35\%$ test EER under mean pooling and $8.76\%$ under weighted pooling, while the proposed PPG~+~XLS-R cross-attention model attains $8.78\%$ and $9.83\%$ respectively. Two characteristics distinguish the results from those obtained for ASVspoof~2019. First, absolute EERs are higher across all configurations, which is the expected consequence of the in-the-wild evaluation conditions: the dataset's crowdsourced acoustic variability and codec-induced degradations introduce a noise floor of difficulty that even the strongest single-stream baselines cannot fully resolve. Second, all four configurations now operate in a narrow Eval EER band of roughly $8\%$--$10\%$. Within this band, the proposed cross-attention model with mean pooling tracks the strongest single-stream baseline closely (within about $0.4$ percentage points), and the residual spread across configurations is small relative to the absolute difficulty of the dataset. The slightly higher EER of the weighted-pooling variant under these conditions is consistent with the substantially greater training-set heterogeneity of ASVspoof~5 that has many more speakers, a wider range of contemporary attacks, and codec-induced variability that does not align cleanly with phonetic-class boundaries, which makes the learned $\alpha_m$ pattern harder to calibrate than on the more homogeneous ASVspoof~2019 partition. We retain weighted pooling for further analyses because the interpretability of the $\alpha_m$ rankings remains the central asset of the framework.

\begin{table}[h!]
\centering
\caption{Detection performance on the ASVspoof~5 Track~1 evaluation partition across the XLS-R-only self-attention baseline and the proposed PPG~+~XLS-R cross-attention model, under mean and weighted pooling. Confidence intervals are shown as \emph{[lower, upper]} values.}
\label{tab:asvs5_results}

\scriptsize
\setlength{\tabcolsep}{4pt}

\resizebox{\linewidth}{!}{%
\begin{tabular}{l l l c c c c}
\hline
Features & Model & Pooling 
& \makecell{Dev EER {[CI]\%}$\downarrow$}
& \makecell{Dev minDCF {[CI]}$\downarrow$}
& \makecell{Eval EER {[CI]\%}$\downarrow$}
& \makecell{Eval minDCF {[CI]}$\downarrow$}
 \\
\hline

\multirow{2}{*}{XLSR}
& \multirow{2}{*}{Self-Attention}
& Mean
& 1.08 [1.01, 1.13]
& 0.08 [0.07, 0.09]
& 8.35 [8.27, 8.43]
& 0.65 [0.64, 0.66]
 \\

& & Weighted
& 1.14 [1.06, 1.20]
& 0.10 [0.09, 0.11]
& 8.76 [8.65, 8.85]
& 0.77 [0.76, 0.79]
 \\

\hline

\multirow{2}{*}{PPG + XLSR}
& \multirow{2}{*}{Cross-Attention}
& Mean
& 1.44 [1.36, 1.52]
& 0.12 [0.11, 0.13]
& 8.78 [8.68, 8.86]
& 0.85 [0.83, 0.87]
 \\

& & Weighted
& 1.99 [1.70, 1.85]
& 0.10 [0.09, 0.11]
& 9.83 [9.74, 9.93]
& 0.77 [0.76, 0.78]
 \\

\hline
\end{tabular}%
}

\end{table}











\begin{figure}[h!]
    \centering
    \includegraphics[width=\linewidth]{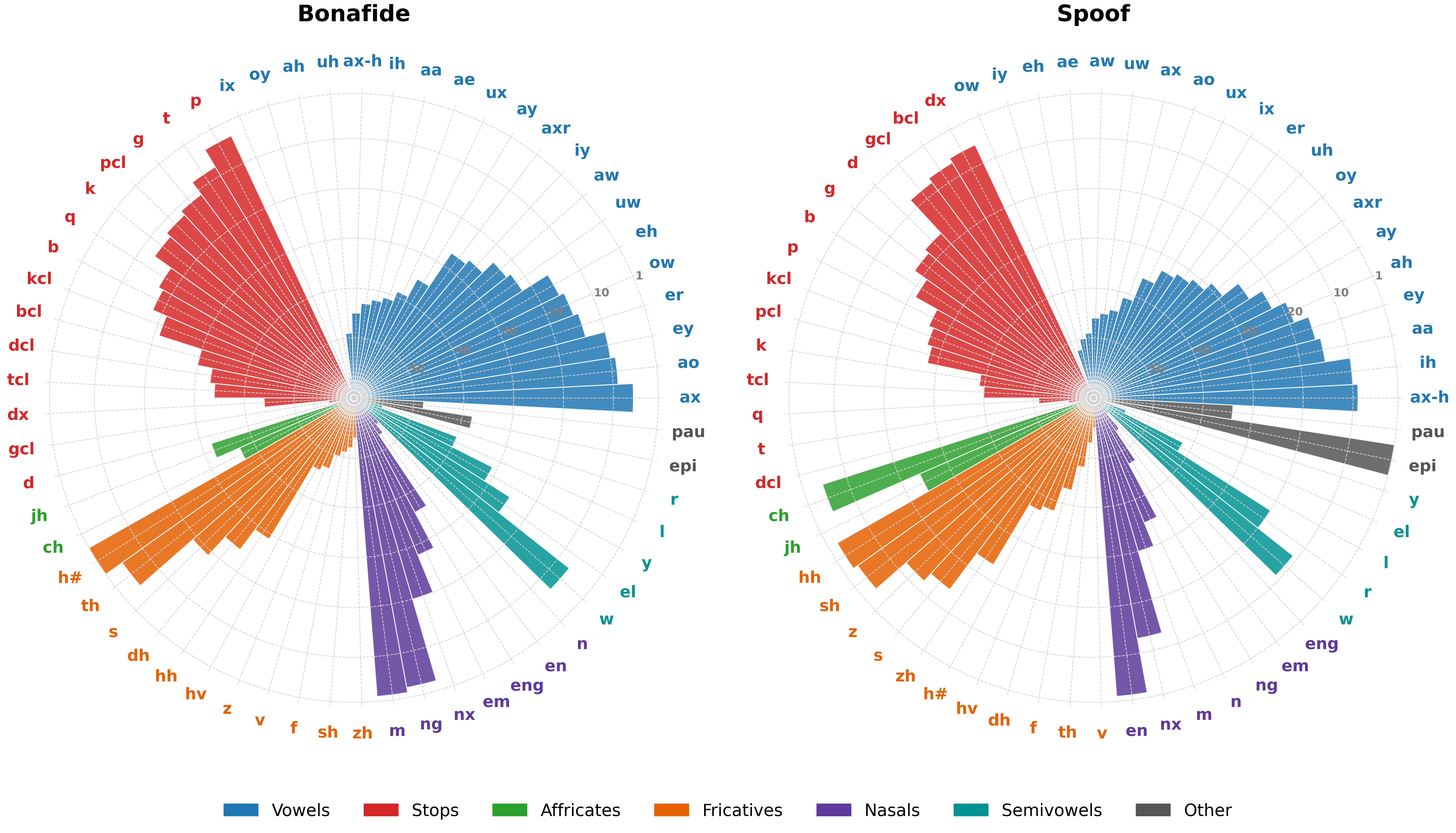}
    \caption{Per-phone importance rankings of the proposed cross-attention model with weighted pooling on ASVspoof 5, derived by ranking the learned weights $\alpha_m$ per utterance and averaging across the bonafide (left) and spoof (right) classes. Bar length encodes average importance: the longest bars correspond to the most strongly attended phones, with stops, fricatives, affrictes and nasals consistently dominating the top-ranked positions across both classes.}
    
    \label{fig:asv5_figure}
\end{figure}

The phone-importance pattern produced by the weighted cross-attention model on ASVspoof~5 is shown in Figure~\ref{fig:asv5_figure}, with the same construction as before. Three observations stand immediately out.

\begin{enumerate}
\item The bonafide and spoof rankings are now \emph{substantially different} for nearly every phone. Whereas on ASVspoof 2019 almost all phones shared the same rank between the two classes, on ASVspoof~5 only a handful of phones occupy similar positions, and several phones move from near the top of one class's ranking to near the bottom of the other (e.g., \texttt{epi}, \texttt{h\#}, \texttt{en}). This indicates that on ASVspoof~5 the bonafide and spoof utterances drive the cross-attention block toward substantially different phone-importance patterns, in contrast to ASVspoof~2019, where both classes induce a single shared ordering and discrimination must rely on the per-phone content of $A_{\text{out}}[m]$;

\item The silence- and closure-related phones (\texttt{pau}, \texttt{h\#}, \texttt{epi}, and the closure variants \texttt{bcl}, \texttt{dcl}, \texttt{gcl}, \texttt{kcl}, \texttt{pcl}, \texttt{tcl}) play a particularly prominent role: \texttt{h\#} sits at the top of the bonafide ranking, while \texttt{epi} sits at the top of the spoof ranking. This finding is notable because our preprocessing pipeline applies energy-based SAD to trim ambient silence \emph{at the beginning and end of each utterance}, while preserving inter-phone closures and brief intra-utterance silences as part of the active speech sequence. The prominence of silence-related phones in the rankings therefore may reflect characteristics of these retained intra-utterance silences rather than the gross silence-shortcut behavior documented elsewhere in the anti-spoofing literature~\cite{chettri2020dataset,muller2021speech}. A more definitive characterization of why these specific phones rank highly under in-the-wild conditions would require a dedicated investigation, which is beyond the scope of the present study;

\item Nasals, fricatives, and several stops consonants continue to feature heavily in the top-ranked positions across both classes (\texttt{m}, \texttt{ng}, \texttt{en}, \texttt{p}, \texttt{th} on the bonafide side; \texttt{en}, \texttt{hh}, \texttt{sh}, \texttt{th}, \texttt{dx}, \texttt{p} on the spoof side), confirming that the obstruent, turbulence-related, and anti-resonance groups identified earlier remain discriminative under in-the-wild conditions, while affricates (\texttt{ch}, \texttt{jh}) emerge as strongly attended specifically on the spoof side.
\end{enumerate}

Taken together, the ASVspoof~5 rankings reinforce that phonetic-group identity is a first-order determinant of detection-relevant information, while showing that the precise mapping from phonetic group to discriminative weight is sensitive to the acoustic conditions and attack diversity of the evaluation set.

\section{Cross-Dataset Synthesis and Interpretable Score Decomposition}
\label{subsec:results_synthesis}

Taken across the three datasets and the targeted ablation study, a coherent and instructive picture emerges. The most important takeaway is that, at least in the conditions tested, discriminative power in deepfake detection does not arise from frequency of occurrence — the phonetic categories that appear most often in speech (vowels and semivowels) are not the categories that carry the strongest detection signal. Instead, discriminative power concentrates on the articulatory categories that present a production mismatch between human speakers and the generator models. Stops, fricatives, affricates, and nasals all involve mechanisms (transient bursts, broadband turbulence, anti-resonance, sound-boundary discontinuities) that are inherently difficult for generators to reproduce at the time-resolution and spectral fidelity that natural speech demands. Vowels and semivowels, by contrast, are produced through periodic, formant-driven mechanisms that modern neural vocoders handle relatively well. The recurring pattern across all three datasets and the ablation study therefore points to a stable observation about how detection works in practice: it is the rarity and complexity of production mechanisms that matters, not the total amount of data.

Beyond this fundamental observation, several practical implications follow. First, the framework's structural transparency makes it possible to audit detection decisions on a per-utterance basis. We argue this could particularly relevant to forensic settings where the reasoning behind a verdict must itself be explainable and defensible. Second, the targeted-masking mechanism provides a built-in tool for understanding which articulatory categories the model is relying on for a given prediction, opening a path for diagnostic workflows that go beyond a single utterance-level score. Third, the dataset-sensitive variation we observe (LJSpeech's strong commitment to stops, ASVspoof~2019's convergence on a class-agnostic ordering, ASVspoof~5's silence-boundary emphasis) provides actionable guidance for practitioners: when deploying detection in different conditions (controlled studio, multi-attack benchmark, in-the-wild codec-degraded audio), one should expect the model's preferred phonetic regions to shift, and the framework's per-utterance interpretability makes it possible to identify which categories are most relied upon under the specific conditions of the deployment. More fundamentally, the convergence of these empirical observations supporting the theoretical assumptions from Section~\ref{section_4} confirms a broader principle for the field: future detection architectures may benefit from explicit phonetic awareness, since the discriminative signal is genuinely concentrated in linguistically meaningful regions, not in undifferentiated frame-level statistics. The interpretive payoff of the framework is therefore not only that it can explain individual decisions, but that it provides empirical evidence for what kind of signal the field should be paying attention to.

While the EER and minDCF numbers reported across the three datasets establish that the framework is detection-competitive with one of the strongest single-stream baselines, the framework's primary contribution lies in its structural interpretability: not just in producing a single utterance-level score, but in exposing the per-phone evidence the model used to arrive at it. Section~\ref{section_4} formalized this interpretability by decomposing the spoofing posterior $P(Y \mid X, W)$ into a weighted sum of phone-conditional evidence terms $P(Y \mid X, Z = z_i)$ modulated by phone-presence weights $w_i$ (Eq.~\eqref{eq:final_factorization}), and the architecture of Section~\ref{arch} realizes both quantities explicitly: the per-row content of $A_{\text{out}}[m]$ supplies the phone-conditional evidence, and the softmax-normalized pooling weights $\alpha_m$ supply the phone-presence weights. Because both quantities are computed on every utterance, the framework allows us to inspect, on a per-utterance basis, \emph{which articulatory groups contributed how much to the final decision} — a property that is structurally unavailable from the single-stream baselines reported earlier.

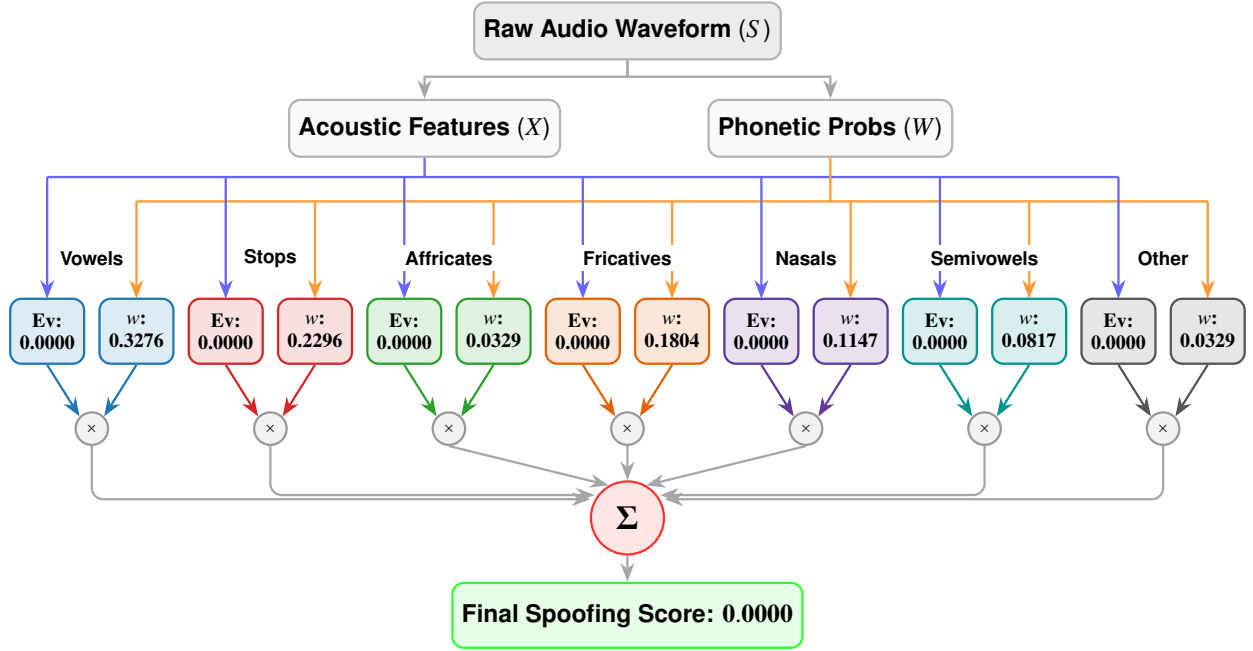
\begin{figure*}[h!]
\centering
\definecolor{colVow}{HTML}{1F77B4}
\definecolor{colStp}{HTML}{D62728}
\definecolor{colAff}{HTML}{2CA02C}
\definecolor{colFrc}{HTML}{E66100}
\definecolor{colNas}{HTML}{5D3A9B}
\definecolor{colSem}{HTML}{009392}
\definecolor{colOth}{HTML}{555555}

\resizebox{\textwidth}{!}{%
\begin{tikzpicture}[
    >=Stealth,
    font=\sffamily\small,
    box/.style={rectangle, draw=gray!60, thick, rounded corners, align=center, minimum height=0.7cm, fill=gray!5},
    gbox/.style={rectangle, draw=#1, thick, rounded corners, align=center, fill=#1!15, minimum width=0.9cm, minimum height=0.8cm, font=\scriptsize\bfseries},
    sum_node/.style={circle, draw=red!80, thick, fill=red!10, minimum size=0.9cm, align=center, inner sep=0pt, font=\large},
    mul_node/.style={circle, draw=gray!80, thick, fill=gray!10, minimum size=0.4cm, align=center, inner sep=0pt, font=\tiny},
    group_label/.style={font=\sffamily\scriptsize\bfseries, align=center, fill=white, inner sep=2pt, text=black},
    arrow/.style={->, thick, draw=gray!70},
    blue_arrow/.style={->, thick, draw=blue!60},
    orange_arrow/.style={->, thick, draw=orange!80},
    blue_line/.style={thick, draw=blue!60},
    orange_line/.style={thick, draw=orange!80}
]

\node[box, fill=gray!15] (audio) at (0, 0) {\textbf{Raw Audio Waveform} ($S$)};
\node[box] (xlsr) at (-2.5, -1.2) {\textbf{Acoustic Features} ($X$)};
\node[box] (ppg) at (2.5, -1.2) {\textbf{Phonetic Probs} ($W$)};

\draw[arrow] (audio.south) -- ++(0,-0.2) -| (xlsr.north);
\draw[arrow] (audio.south) -- ++(0,-0.2) -| (ppg.north);

\def\boxSep{0.55} 
\def\gA{-6.6} \def\gB{-4.4} \def\gC{-2.2} \def\gD{0} \def\gE{2.2} \def\gF{4.4} \def\gG{6.6}

\draw[blue_line] (xlsr.south) -- (-2.5, -1.8);
\draw[blue_line] (\gA - \boxSep, -1.8) -- (\gG - \boxSep, -1.8);

\draw[orange_line] (ppg.south) -- (2.5, -2.1);
\draw[orange_line] (\gA + \boxSep, -2.1) -- (\gG + \boxSep, -2.1);

\foreach \x in {\gA, \gB, \gC, \gD, \gE, \gF, \gG} {
    \draw[blue_arrow] (\x - \boxSep, -1.8) -- (\x - \boxSep, -3.3);
    \draw[orange_arrow] (\x + \boxSep, -2.1) -- (\x + \boxSep, -3.3);
}

\node[group_label] (l_vowel) at (\gA, -2.8) {Vowels};
\node[gbox=colVow] (e_vowel) at (\gA - \boxSep, -3.7) {Ev:\\0.0000};
\node[gbox=colVow] (w_vowel) at (\gA + \boxSep, -3.7) {$w$:\\0.3276};
\node[mul_node] (mul1) at (\gA, -4.9) {$\times$};

\node[group_label] (l_stop) at (\gB, -2.8) {Stops};
\node[gbox=colStp] (e_stop) at (\gB - \boxSep, -3.7) {Ev:\\0.0000};
\node[gbox=colStp] (w_stop) at (\gB + \boxSep, -3.7) {$w$:\\0.2296};
\node[mul_node] (mul2) at (\gB, -4.9) {$\times$};

\node[group_label] (l_affric) at (\gC, -2.8) {Affricates};
\node[gbox=colAff] (e_affric) at (\gC - \boxSep, -3.7) {Ev:\\0.0000};
\node[gbox=colAff] (w_affric) at (\gC + \boxSep, -3.7) {$w$:\\0.0329};
\node[mul_node] (mul3) at (\gC, -4.9) {$\times$};

\node[group_label] (l_fric) at (\gD, -2.8) {Fricatives};
\node[gbox=colFrc] (e_fric) at (\gD - \boxSep, -3.7) {Ev:\\0.0000};
\node[gbox=colFrc] (w_fric) at (\gD + \boxSep, -3.7) {$w$:\\0.1804};
\node[mul_node] (mul4) at (\gD, -4.9) {$\times$};

\node[group_label] (l_nasal) at (\gE, -2.8) {Nasals};
\node[gbox=colNas] (e_nasal) at (\gE - \boxSep, -3.7) {Ev:\\0.0000};
\node[gbox=colNas] (w_nasal) at (\gE + \boxSep, -3.7) {$w$:\\0.1147};
\node[mul_node] (mul5) at (\gE, -4.9) {$\times$};

\node[group_label] (l_semi) at (\gF, -2.8) {Semivowels};
\node[gbox=colSem] (e_semi) at (\gF - \boxSep, -3.7) {Ev:\\0.0000};
\node[gbox=colSem] (w_semi) at (\gF + \boxSep, -3.7) {$w$:\\0.0817};
\node[mul_node] (mul6) at (\gF, -4.9) {$\times$};

\node[group_label] (l_other) at (\gG, -2.8) {Other};
\node[gbox=colOth] (e_other) at (\gG - \boxSep, -3.7) {Ev:\\0.0000};
\node[gbox=colOth] (w_other) at (\gG + \boxSep, -3.7) {$w$:\\0.0329};
\node[mul_node] (mul7) at (\gG, -4.9) {$\times$};

\foreach \e/\w/\m/\c in {
    e_vowel/w_vowel/mul1/colVow,
    e_stop/w_stop/mul2/colStp,
    e_affric/w_affric/mul3/colAff,
    e_fric/w_fric/mul4/colFrc,
    e_nasal/w_nasal/mul5/colNas,
    e_semi/w_semi/mul6/colSem,
    e_other/w_other/mul7/colOth} {
    \draw[->, thick, draw=\c] (\e.south) -- (\m.north west);
    \draw[->, thick, draw=\c] (\w.south) -- (\m.north east);
}

\node[sum_node] (sum) at (0, -6.0) {$\boldsymbol{\Sigma}$};

\draw[arrow, rounded corners=4pt] (mul1.south) |- ([yshift=0.15cm]sum.170);
\draw[arrow, rounded corners=4pt] (mul2.south) |- ([yshift=0.05cm]sum.150);
\draw[arrow, rounded corners=4pt] (mul3.south) -- (sum.120);
\draw[arrow] (mul4.south) -- (sum.north);
\draw[arrow, rounded corners=4pt] (mul5.south) -- (sum.60);
\draw[arrow, rounded corners=4pt] (mul6.south) |- ([yshift=0.05cm]sum.30);
\draw[arrow, rounded corners=4pt] (mul7.south) |- ([yshift=0.15cm]sum.10);

\node[box, fill=green!10, draw=green!80, thick, minimum height=0.8cm] (output) at (0, -7.2) {\textbf{Final Spoofing Score:} $\mathbf{0.0000}$};
\draw[arrow] (sum.south) -- (output.north);

\end{tikzpicture}
}
\caption{Diagrammatic representation of the score decomposition for a bonafide utterance (\texttt{E\_0000000192.flac}) from the ASVspoof 5 evaluation set. The model detects a complete absence of spoofing artifacts across all seven phonetic categories (evidence values of $0.0000$), uniformly contributing to an accurate and definitive bonafide classification.}
\label{fig:decomp_bonafide}
\end{figure*}

\begin{figure*}[h!]
\centering
\definecolor{colVow}{HTML}{1F77B4}
\definecolor{colStp}{HTML}{D62728}
\definecolor{colAff}{HTML}{2CA02C}
\definecolor{colFrc}{HTML}{E66100}
\definecolor{colNas}{HTML}{5D3A9B}
\definecolor{colSem}{HTML}{009392}
\definecolor{colOth}{HTML}{555555}

\resizebox{\textwidth}{!}{%
\begin{tikzpicture}[
    >=Stealth,
    font=\sffamily\small,
    box/.style={rectangle, draw=gray!60, thick, rounded corners, align=center, minimum height=0.7cm, fill=gray!5},
    gbox/.style={rectangle, draw=#1, thick, rounded corners, align=center, fill=#1!15, minimum width=0.9cm, minimum height=0.8cm, font=\scriptsize\bfseries},
    sum_node/.style={circle, draw=red!80, thick, fill=red!10, minimum size=0.9cm, align=center, inner sep=0pt, font=\large},
    mul_node/.style={circle, draw=gray!80, thick, fill=gray!10, minimum size=0.4cm, align=center, inner sep=0pt, font=\tiny},
    group_label/.style={font=\sffamily\scriptsize\bfseries, align=center, fill=white, inner sep=2pt, text=black},
    blue_arrow/.style={->, thick, draw=blue!60},
    arrow/.style={->, thick, draw=gray!70},
    red_arrow/.style={->, thick, draw=red!60},
    orange_arrow/.style={->, thick, draw=orange!80},
    red_line/.style={thick, draw=red!60},
    blue_line/.style={thick, draw=blue!60},
    orange_line/.style={thick, draw=orange!80}
]

\node[box, fill=gray!15] (audio) at (0, 0) {\textbf{Raw Audio Waveform} ($S$)};
\node[box] (xlsr) at (-2.5, -1.2) {\textbf{Acoustic Features} ($X$)};
\node[box] (ppg) at (2.5, -1.2) {\textbf{Phonetic Probs} ($W$)};

\draw[arrow] (audio.south) -- ++(0,-0.2) -| (xlsr.north);
\draw[arrow] (audio.south) -- ++(0,-0.2) -| (ppg.north);

\def\boxSep{0.55}
\def\gA{-6.6} \def\gB{-4.4} \def\gC{-2.2} \def\gD{0} \def\gE{2.2} \def\gF{4.4} \def\gG{6.6}

\draw[blue_line] (xlsr.south) -- (-2.5, -1.8);
\draw[blue_line] (\gA - \boxSep, -1.8) -- (\gG - \boxSep, -1.8);

\draw[orange_line] (ppg.south) -- (2.5, -2.1);
\draw[orange_line] (\gA + \boxSep, -2.1) -- (\gG + \boxSep, -2.1);

\foreach \x in {\gA, \gB, \gC, \gD, \gE, \gF, \gG} {
    \draw[blue_arrow] (\x - \boxSep, -1.8) -- (\x - \boxSep, -3.3);
    \draw[orange_arrow] (\x + \boxSep, -2.1) -- (\x + \boxSep, -3.3);
}

\node[group_label] (l_vowel) at (\gA, -2.8) {Vowels};
\node[gbox=colVow] (e_vowel) at (\gA - \boxSep, -3.7) {Ev:\\0.9577};
\node[gbox=colVow] (w_vowel) at (\gA + \boxSep, -3.7) {$w$:\\0.3256};
\node[mul_node] (mul1) at (\gA, -4.9) {$\times$};

\node[group_label] (l_stop) at (\gB, -2.8) {Stops};
\node[gbox=colStp] (e_stop) at (\gB - \boxSep, -3.7) {Ev:\\0.9566};
\node[gbox=colStp] (w_stop) at (\gB + \boxSep, -3.7) {$w$:\\0.2302};
\node[mul_node] (mul2) at (\gB, -4.9) {$\times$};

\node[group_label] (l_affric) at (\gC, -2.8) {Affricates};
\node[gbox=colAff] (e_affric) at (\gC - \boxSep, -3.7) {Ev:\\0.9528};
\node[gbox=colAff] (w_affric) at (\gC + \boxSep, -3.7) {$w$:\\0.0341};
\node[mul_node] (mul3) at (\gC, -4.9) {$\times$};

\node[group_label] (l_fric) at (\gD, -2.8) {Fricatives};
\node[gbox=colFrc] (e_fric) at (\gD - \boxSep, -3.7) {Ev:\\0.9563};
\node[gbox=colFrc] (w_fric) at (\gD + \boxSep, -3.7) {$w$:\\0.1814};
\node[mul_node] (mul4) at (\gD, -4.9) {$\times$};

\node[group_label] (l_nasal) at (\gE, -2.8) {Nasals};
\node[gbox=colNas] (e_nasal) at (\gE - \boxSep, -3.7) {Ev:\\0.9572};
\node[gbox=colNas] (w_nasal) at (\gE + \boxSep, -3.7) {$w$:\\0.1144};
\node[mul_node] (mul5) at (\gE, -4.9) {$\times$};

\node[group_label] (l_semi) at (\gF, -2.8) {Semivowels};
\node[gbox=colSem] (e_semi) at (\gF - \boxSep, -3.7) {Ev:\\0.9594};
\node[gbox=colSem] (w_semi) at (\gF + \boxSep, -3.7) {$w$:\\0.0800};
\node[mul_node] (mul6) at (\gF, -4.9) {$\times$};

\node[group_label] (l_other) at (\gG, -2.8) {Other};
\node[gbox=colOth] (e_other) at (\gG - \boxSep, -3.7) {Ev:\\0.9522};
\node[gbox=colOth] (w_other) at (\gG + \boxSep, -3.7) {$w$:\\0.0342};
\node[mul_node] (mul7) at (\gG, -4.9) {$\times$};

\foreach \e/\w/\m/\c in {
    e_vowel/w_vowel/mul1/colVow,
    e_stop/w_stop/mul2/colStp,
    e_affric/w_affric/mul3/colAff,
    e_fric/w_fric/mul4/colFrc,
    e_nasal/w_nasal/mul5/colNas,
    e_semi/w_semi/mul6/colSem,
    e_other/w_other/mul7/colOth} {
    \draw[->, thick, draw=\c] (\e.south) -- (\m.north west);
    \draw[->, thick, draw=\c] (\w.south) -- (\m.north east);
}

\node[sum_node] (sum) at (0, -6.0) {$\boldsymbol{\Sigma}$};

\draw[arrow, rounded corners=4pt] (mul1.south) |- ([yshift=0.15cm]sum.170);
\draw[arrow, rounded corners=4pt] (mul2.south) |- ([yshift=0.05cm]sum.150);
\draw[arrow, rounded corners=4pt] (mul3.south) -- (sum.120);
\draw[arrow] (mul4.south) -- (sum.north);
\draw[arrow, rounded corners=4pt] (mul5.south) -- (sum.60);
\draw[arrow, rounded corners=4pt] (mul6.south) |- ([yshift=0.05cm]sum.30);
\draw[arrow, rounded corners=4pt] (mul7.south) |- ([yshift=0.15cm]sum.10);

\node[box, fill=red!10, draw=red!80, thick, minimum height=0.8cm] (output) at (0, -7.2) {\textbf{Final Spoofing Score:} $\mathbf{0.9569}$};
\draw[arrow] (sum.south) -- (output.north);

\end{tikzpicture}
}
\caption{Diagrammatic representation of the score decomposition for a spoof utterance (\texttt{E\_0000000034.flac}) from the ASVspoof 5 evaluation set. The model identifies strong acoustic indicators of spoofing (evidence values $> 0.95$) uniformly across every phonetic category, resulting in an accurate and robust overall spoof classification.}
\label{fig:decomp_spoof}
\end{figure*}
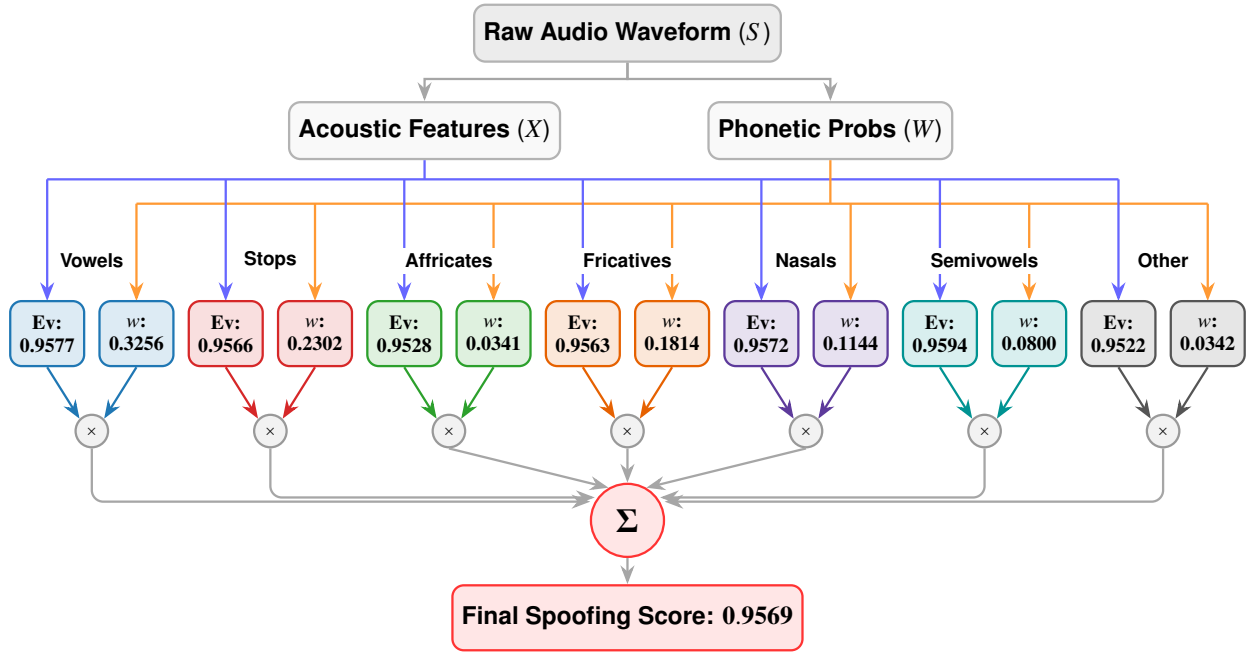

We illustrate this interpretability through two case studies drawn from the ASVspoof~5 evaluation set. Figure~\ref{fig:decomp_bonafide} shows the score decomposition for a bonafide utterance (\texttt{E\_0000000192.flac}), and Figure~\ref{fig:decomp_spoof} shows the decomposition for a spoof utterance (\texttt{E\_0000000034.flac}). In both diagrams, the per-group phone-conditional evidence values $P(Y \mid X, Z = z_g)$ (blue) and the corresponding group-level phone-presence weights $w_g$ (orange), aggregated over the seven articulatory groups of Table~\ref{tab:phoneme_groups}, are reported alongside the weighted-sum aggregation that produces the final score. The two case studies illustrate the complementary qualitative regimes that the framework's interpretability is designed to surface. On the bonafide utterance, the model produces near-zero phone-conditional evidence values across \emph{every} articulatory group, with each group contributing uniformly to a definitive bonafide classification (final score $\approx 0$). On the spoof utterance, the model produces high phone-conditional evidence values (all above $0.95$) across \emph{every} articulatory group, with each group contributing uniformly to a definitive spoof classification (final score $\approx 0.96$). Critically, both decisions are not just delivered as numerical scores but accompanied by an inspectable per-group breakdown that exposes the contribution of each articulatory category to the final verdict, which is precisely the architectural property argued for in Sections~\ref{section_4} and~\ref{subsec:backend}.

\section{Conclusion}

Our work puts forward a phoneme-guided cross-attention framework for interpretable, phonetically-grounded speech deepfake detection. By probabilistically factorizing the detection objective, our model explicitly decomposes an utterance's authenticity score into phone-conditional acoustic evidence modulated by continuous phonetic presence weights. Across our experiments, we demonstrated that fusing self-supervised acoustic representations (XLS-R) with phonetic posteriorgrams (PPGs) via cross-attention yields an effective structural scaffold for localizing synthesis artifacts. Our extensive evaluation on three distinct datasets presenting both legacy and state-of-the-art speech deepfakes revealed consistent patterns in how generative pipelines fail across different articulatory categories, supported by strong empirical performance from our proposed weighted cross-attention model. 

While our work serves as a strong foundation for phonetically-explainable anti-spoofing, several broader factors relevant to forensic casework and real-world deployment deserve their own dedicated follow-up studies. These include robustness under coding and transcoding artifacts, additive noise, and other source-degradation conditions that arise outside controlled benchmark settings. Each additional nuisance variation is expected to interact with fine-grained phonetic structure in ways the present framework is not designed to characterize but is well-positioned to investigate. Cross-language evaluation is another natural extension: by exposing exactly which phonetic categories drive each decision, the framework provides a transparent diagnostic for understanding how detection performance generalizes (or fails to generalize) across languages with different phonetic inventories. Finally, from a computational standpoint, the dual-extractor design (XLS-R for acoustic features and a fine-tuned Wav2Vec 2.0 for phonetic posteriorgrams) introduces additional parameter footprint compared to single-stream detectors. While both extractors are frozen and pre-computed offline in our current implementation, exploring more economical formulations, for example, a shared backbone with multi-task heads producing both acoustic and phonetic representations, represents a promising direction for reducing the overall computational burden without compromising the framework's interpretive properties.

\section*{Acknowledgments}
The work has been partially supported by the Academy of Finland (Decision No. 349605, project ``SPEECHFAKES''). The authors wish to acknowledge CSC—IT Center for Science, Finland, for computational resources.

\appendix

\section{Empirical Justification of Assumption 3: Uninformative Phonetic Priors}
\label{app:assumption3}

Assumption~3 of our probabilistic factorization (Section~\ref{section_4}) states that
\begin{equation}
P(Y \mid Z=z_i) = P(Y),
\end{equation}
i.e., the latent phonetic class $Z$ provides no inherent evidence about audio authenticity; equivalently, $Y$ and $Z$ are statistically independent, written $Y \perp\!\!\!\perp Z$. To address the adequacy of this assumption empirically, we extract frame-level phone posteriors from every utterance in the ASVspoof~2019 LA training partition and estimate the soft phonetic prior for each class as
\begin{equation}
\hat{P}(Z=z_i \mid Y=y)
=
\frac{1}{N_y}
\sum_{u \in \mathcal{U}_y}
\sum_{t=1}^{T_u}
w_{i,t}^{(u)},
\end{equation}
where $\mathcal{U}_y$ is the set of utterances with class label $y \in \{0,1\}$ (spoofed or bonafide), $T_u$ is the number of frames in utterance $u$, $w_{i,t}^{(u)}$ is the soft phonetic posterior $P(Z=z_i \mid S)$ at frame $t$ of utterance $u$, and
\begin{equation}
N_y
=
\sum_{u \in \mathcal{U}_y}
T_u
\end{equation}
is the total number of frames in class $y$.

\begin{figure}[h!]
\centering
\includegraphics[width=\linewidth]{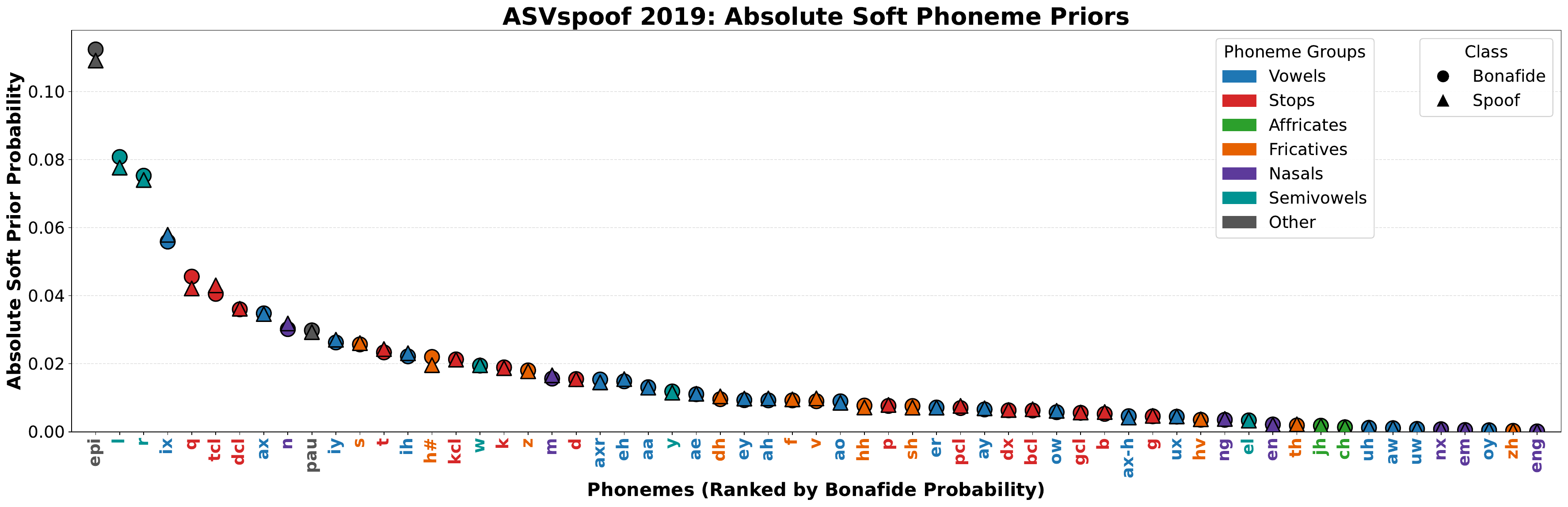}
\caption{Per-phone soft priors $\hat{P}(Z=z_i \mid Y=y)$ for bonafide versus spoof on the ASVspoof~2019 LA training partition. All 61 phones lie tightly along the identity line $y=x$, indicating that the bonafide and spoof phonetic distributions are nearly identical; the most frequent phones (visible in the upper-left) and the rarer phones (clustered near the origin) all exhibit the same agreement.}
\label{fig:prior_scatter}
\end{figure}

\begin{figure}[h!]
\centering
\includegraphics[width=\linewidth]{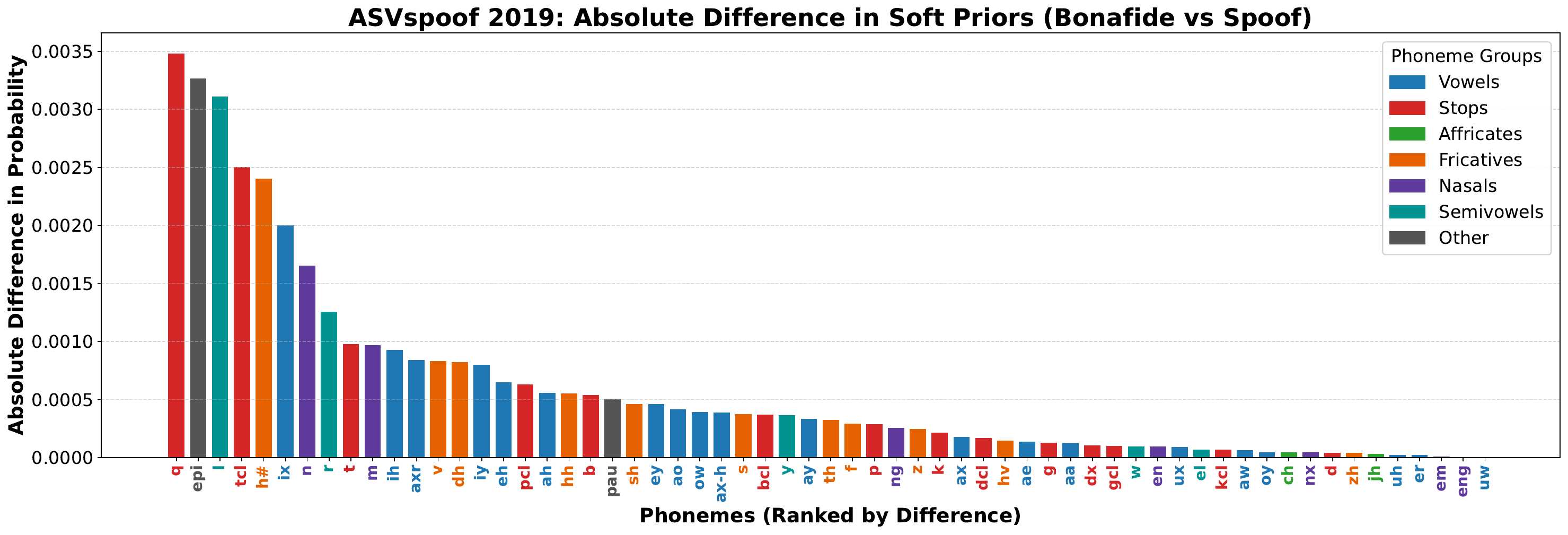}
\caption{Per-phone absolute difference between bonafide and spoof soft priors. The maximum discrepancy across all 61 phones is $\approx 3.5 \times 10^{-3}$.}
\label{fig:prior_diff_bar}
\end{figure}

Figure~\ref{fig:prior_scatter} plots the per-phone priors for each of the 61 TIMIT phones across the bonafide and spoof classes. Figure~\ref{fig:prior_diff_bar} quantifies the residual gap: the maximum absolute difference across the entire inventory is approximately $3.5 \times 10^{-3}$, and across all 61 phones, the absolute differences cluster near zero. The phones with the largest residual differences (\texttt{q}, \texttt{epi}, \texttt{tcl}, \texttt{h\#}) are silence- and closure-related markers, which are particularly susceptible to small sample-size fluctuations because their phonetic boundaries are inherently noisier and more dependent on the specific recording conditions of each utterance. The overall pattern of vanishingly small differences across the entire phonetic inventory, mirrors one of the design objectives of speech synthesis systems: to maintain intelligibility, deepfake generators are trained to reproduce the same linguistic content as authentic speech, so any systematic over- or under-representation of particular phones would be a trivially exploitable artifact and is explicitly avoided. The marginal $P(Z)$ therefore carries essentially no information about $Y$, justifying the simplification in Assumption~3.

A complementary observation strengthens this argument: when these priors are compared with the phone-importance ranking produced by our trained model (Figure~\ref{fig:ASVS19}), the most attended phones are neither the most frequent nor those with the largest absolute prior differences---\texttt{epi}, despite topping both rankings, does not appear among the top-attended phones, while several phones with negligible prior differences receive substantial attention. The model's discriminative signal therefore comes from \emph{how} a phone is acoustically realized, not from \emph{which} phones are uttered.

\section{Loss Convergence on ASVspoof~2019}
\label{app:convergence}

\begin{figure}[h!]
\centering
\includegraphics[width=0.85\linewidth]{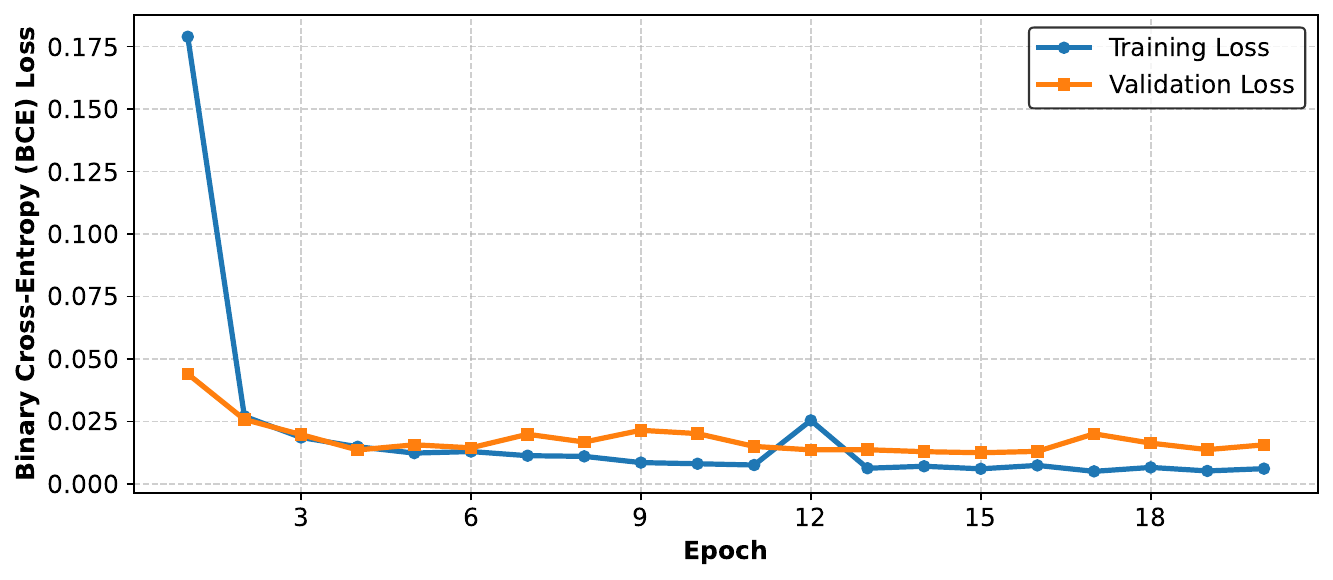}
\caption{Training and validation BCE loss for the proposed cross-attention model (weighted pooling) on ASVspoof~2019 LA, plotted over the 20-epoch training budget. Both curves flatten well within the budget, confirming that 20 epochs is a sufficient stopping point for the converged regime.}
\label{fig:loss_curves}
\end{figure}

Figure~\ref{fig:loss_curves} shows the training and validation binary cross-entropy (BCE) loss of the proposed cross-attention model with weighted pooling on the ASVspoof~2019 LA partition over the 20 epochs used in our experiments. Both curves descend rapidly in the first few epochs and largely flatten well before the 20-epoch cap, supporting the choice of this training budget as a reasonable compromise between computational cost and loss convergence. The selected checkpoint corresponds to the epoch of lowest validation EER (Table~\ref{tab:asvs19_table}).

 \bibliographystyle{elsarticle-num} 
 \bibliography{refs}
\end{document}